\documentclass[11pt,natbib]{article}

\usepackage{geometry}
\usepackage{authblk}
\usepackage{amsmath}
\usepackage{amssymb}
\usepackage{dsfont}
\usepackage{url}
\usepackage{subcaption}
\usepackage{mathrsfs}
\usepackage{float} 
\usepackage{graphicx}
\allowdisplaybreaks
\usepackage{bibunits}
\usepackage{csquotes}
\usepackage{minitoc}
\usepackage{tocloft}
\usepackage{enumitem}
\cftsetindents{sec}{0em}{2.75em}
\cftsetindents{subsec}{2.75em}{3.5em}

\usepackage{titletoc}

\usepackage{amsthm}

\graphicspath{{./Figures/}}

\theoremstyle{definition}
\makeatletter
\newtheorem*{rep@theorem}{\rep@title}
\newcommand{\newreptheorem}[2]{%
\newenvironment{rep#1}[1]{%
 \def\rep@title{#2 \ref{##1}}%
 \begin{rep@theorem}}%
 {\end{rep@theorem}}}
\makeatother

\newtheorem{theorem}{Theorem}
\newreptheorem{theorem}{Theorem}

\newreptheorem{lemma}{Lemma}
\newtheorem*{theorem*}{Theorem}

\newtheorem{corollary}{Corollary}
\newreptheorem{corollary}{Corollary}

\newtheorem{assumption}{Assumption}

\usepackage[ruled,linesnumbered]{algorithm2e}

\usepackage{natbib}

\newcommand*{\defeq}{\stackrel{\text{def}}{=}}

\makeatletter
\newcommand*{\indep}{%
 \mathbin{%
   \mathpalette{\@indep}{}%
 }%
}
\newcommand*{\nindep}{%
 \mathbin{%
   \mathpalette{\@indep}{\not}%
 }%
}
\newcommand*{\@indep}[2]{%
 \sbox0{$#1\perp\m@th$}%
 \sbox2{$#1=$}%
 \sbox4{$#1\vcenter{}$}%
 \rlap{\copy0}%
 \dimen@=\dimexpr\ht2-\ht4-.2pt\relax
 \kern\dimen@
 {#2}%
 \kern\dimen@
 \copy0 %
}
\makeatother

\title{Stacking interventions for equitable outcomes}
\author{James Liley}
\affil{\small Durham University, UK}

\begin{document}
\begin{bibunit}[custom]

\maketitle

\begin{abstract}

Predictive risk scores estimating probabilities for a binary outcome on the basis of observed covariates are common across the sciences. They are frequently developed with the intent of avoiding the outcome in question by intervening in response to estimated risks. 
Since risk scores are usually developed in complex systems, interventions usually take the form of expert agents responding to estimated risks as they best see fit. In this case, interventions may be complex and their effects difficult to observe or infer, meaning that explicit specification of interventions in response to risk scores is impractical. Scope to modulate the aggregate model-intervention scheme so as to optimise an objective is hence limited.
We propose an algorithm by which a model-intervention scheme can be developed by 'stacking' possibly unknown intervention effects. By repeatedly observing and updating the intervention and model, we show that this scheme leads to convergence or almost-convergence of eventual outcome risk to an equivocal value for any initial value of covariates. Our approach deploys a series of risk scores to expert agents, with instructions to act on them in succession according to their best judgement.
Our algorithm uses only observations of pre-intervention covariates and the eventual outcome as input. It is not necessary to know or infer the effect of the intervention, other than making a general assumption that it is 'well-intentioned'. The algorithm can also be used to safely update risk scores in the presence of unknown interventions and concept drift. We demonstrate convergence of expectation of outcome in a range of settings and show robustness to errors in risk estimation and to concept drift.
We suggest several practical applications and demonstrate a potential implementation by simulation, showing that the algorithm leads to a fair distribution of outcome risk across a population.
\end{abstract}








\clearpage

\section{Introduction}
\label{sec:introduction}

Prediction of a binary outcome is a common problem. A typical setting concerns a set of predictors $X \in \mathbb{R}^p$ and an outcome $Y \in \{0,1\}$, with conditional distribution determined by $P(Y|X)=\mathbb{E}(Y|X)=f(X)$. An approximation of $f$ is inferred from training instances of $(X,Y)$, giving a `risk score' $\rho \approx f$, which can then be used on new samples $x$ (for which $Y$ is unknown) to generate predictions $\rho(x)\approx P(Y|X=x)=f(x)$~\citep{friedman01}. 

The practical aim of predicting an outcome is usually to try and avoid it occurring. This may be accomplished by intervening on the covariates $X$ of samples in response to the estimated risk $\rho(X)$. Since settings are typically too complex to safely prescribe specific interventions, this is often achieved by making values of $\rho(X)$ available to expert agents, who may then intervene as they see fit. For instance, in medical settings, we may intervene on predicted risks by reporting risk scores to doctors~\citep{hyland20,artzi20}. We may consider the aggregate of the risk score and set of interventions made in response to it as aiming to minimise incidence of the outcome. 

It is generally straightforward for expert agents to make `well-intentioned' interventions, which we can be confident will move risk down (or up) by some amount. However, the effect of such interventions may be difficult to infer or measure~\citep{komorowski18}. For example, a medical practitioner may respond to an increased predicted risk of coronary events by increasing frequency of follow-up and recommending lifestyle changes. While we can be confident that this will reduce coronary risk, the reduction is effected by nebulous effects on multiple risk factors and is difficult to measure~\citep{hippisley17}. This difficulty means that the calibration of intervention to estimated risk is a complex task~\citep{finlayson20}, especially when cost constraints on interventions necessitate careful distribution of resources.


We propose a simple algorithm (which we call `stacked interventions) which enables calibration of intervention \emph{without} requiring expert agents to change behaviour. Namely, all we require is that experts follow the instruction, given a patient with covariates $X$ and risk score $\rho$:
\begin{displayquote}
Presume that $\rho$ represents the risk of the outcome in sample with covariates $X$ and act accordingly
%
\end{displayquote}
and to do this repeatedly for a series of risk scores $\rho_0$, $\rho_1$, $\rho_2$, $\dots$. The series of scores we will use all genuinely correspond to outcome risks in given circumstances (rather than being allowed to vary arbitrarily), ensuring that experts are never asked to act contrary to their own judgement. We will show that this simple course of action leads to an almost-correct calibration of response and optimal distribution of intervention, the latter in the sense of eventually leading to sample-specific interventions which bring each sample to approximately the same level of risk. We require only that it be possible to observe covariates $X$ before any intervention has taken place, and outcomes $Y$ determined by the values of $X$ after intervention, and that it be possible to `update' risk scores over time after making new such observations. We show that the property of near-convergence is robust to `drift' in the system being modelled, which is a common problem in risk scores~\citep{tahmasbi20,davis19}. It is also robust to inaccuracies in risk estimates, in that errors in successive risk scores $\rho_0$, $\rho_1$, $\rho_2$, $\dots$ do not correspond to accumulating errors in the intervention.

Our approach essentially constitutes a strategy for predicive score updating. As such, it allows a resolution of a paradox which has received recent attention~\citep{sperrin18,lenert19,perdomo20,liley21a,izzo21}. If a predictive risks score is `used' in the sense of guiding interventions to avoid the outcome, then the risk score leads to a change in the distribution it aims to model. Should the risk score be updated, and the new risk score simply replace the old, this can lead to dangerous biases in prediction. For instance, if a patient with covariates $X$ is assessed as high-risk for a coronary event, and in response a doctor intervenes, hence lowering their risk, then when we come to update the risk score it appears that covariates $X$ are now not associated with a high risk. If the new risk score replaces the old, then the new risk score underestimates the patient's risk, potentially dangerously. This problem worsens the more the risk score is used. Since most risk scores need to be updated due to `drift' in the distribution they model~\citep{tahmasbi20,davis19} and are intended to be used to avoid an outcome, this problem is widespread and lacks a convenient resolution~\citep{haidar22}.



A real-world evaluation of our method would necessitate a setting of sufficient seriousness that predictive models are used to avoid bad outcomes, but of sufficient flexibility that a new method (ours) be usable to safely propose interventions. This is impractical at early stages of method development, so the purpose of this work is restricted to development of theory and demonstration in hypothetical settings. After outlining related work (section~\ref{sec:litrev}), we introduce notation and preliminary observations, and describe our main algorithm (section~\ref{sec:setup}). We develop properties of this algorithm (sections~\ref{sec:results}, \ref{sec:robustness}): almost-convergence of risks, tolerance to inaccurate risk scores and robustness to drift. Finally, we demonstrate results by simulation (section~\ref{sec:motivating_example}), and describe several potential applications on real-world predictive scores.

\section{Related work}
\label{sec:litrev}

This work concerns a method to periodically update predictive scores. Repeatedly updating a risk model is a common practice, and generally required due to `drift', a gradual change in the joint distribution of $(X,Y)$~\citep{davis19,tahmasbi20,subbaswamy21}. Two examples are the EuroSCORE project for cardiovascular surgical risk prediction~\citep{roques99,nashef12} and the QRISK project for cardiac event risk~\citep{hippisley07,hippisley08,hippisley17}. In both examples, and typically, a `naive' updating strategy is used, in which covariates (e.g., blood biochemistry, blood pressure) and outcome are observed in the population and used to fit a new predictive model, which replaces the original.

A problem with this naive updating strategy was noted in~\citep{lenert19}, which described prognostic models as becoming `victims of their own success': as in section~\ref{sec:introduction}, if the score leads to agents acting on covariate values to reduce risk, updated risk scores will be biased. Failing to account for this effect can be dangerous, particularly in healthcare settings~\citep{liley21a}. This phenomenon is typically modelled causally~\citep{sperrin18,sperrin19}. 

Perdomo et al consider a setting in which naive updating (which they term `repeated risk minimisation') is done with knowledge of this effect (termed `performative prediction')~\citep{perdomo20}.
Taking the new distribution of $(X,Y)$ after the actions induced by a risk score by parameters $\theta$ as $\mathscr{D}(\theta)$, and considering a loss function  $\ell(X,Y,\theta)$, `performative stability' is defined by a set of parameters $\theta_{PS}$ satisfying
\begin{equation}
\theta_{PS}=\arg\min_{\theta} \mathbb{E}_{(X,Y) \sim \mathscr{D}(\theta_{PS})} \ell\left((X,Y),\theta\right) \nonumber
\end{equation}
If risk score updating proceeds naively, successive predictive scores converge (under reasonable conditions) to performative stability: essentially, they predict incidence of outcomes \emph{after} interventions taken in response to the predictions themselves. 
%
%
%
%
Other algorithms for model updating in which the aim is convergence to performative stability are given in~\citep{mendler20}, ~\citep{drusvyatskiy20}, \citep{izzo21} and~\citep{li21}. 

Performatively stable risk scores do not necessarily optimise distribution of interventions, nor minimise incidence of the outcome, and hence are not universally desirable. If we have two samples at pre-intervention risk $P(Y|X)=80$\%, one of whom can be intervened on to a varying degree to bring risk down to 0\%, the other who cannot, and suppose that interventions are allocated in response to risk score, a performatively stable score would assign a score of (around) 40\% to the interventionable sample, prompting a (lacklustre) intervention which reduces their risk to 40\%, and a score of 80\% to the non-interventionable sample, prompting a heavy intervention with no effect. An alternative (swapped) distribution of interventions would achieve a better result: risks of 0\% and 80\% for the same cost.

One way to avoid performative prediction effects is to jointly model the intervention and score. Alaa et al and Sperrin et al propose causal models to infer the effects of intervention as well as the distribution of $(X,Y)$~\citep{alaa18,sperrin19}. This generally requires more data collection than repeated risk minimisation; we show in section~\ref{sec:preliminaries} that the form of an intervention on covariates is not generally identifiable given only pre-intervention covariate values and post-intervention binary outcome. A second simple option to avoid performative prediction is to use a `held-out' set of samples (on whom no risk score is calculated, and hence no risk-score guided interventions occur) to refit a new risk score~\citep{haidar22}. This has clear ethical drawbacks in that held-out samples receive no benefit from the risk score.

Our algorithm be considered as a reinforcement learning problem with constraints on possible decisions. An approach in~\citep{komorowski18} models a series of clinical decisions as a Markov decision process (MDP), amending planned treatment based on response. Again, this requires observation of covariates (in this case, patient status) before and after interventions are made. In general, reinforcement learning evaluation in healthcare is difficult~\citep{gottesman18} meaning policies robust to deviations from recommended interventions are preferred.

Our aim of reducing incidence of the outcome while being unable to `afford' to maximally intervene universally can be considered a problem of resource allocation. In the context of predictive scores, this has been studied in terms of fairness, or equitable access to treatment over sample subgroups. One relevant algorithm~\citep{elzayn19} uses an iterated algorithm to allow a `distributor' to attain an equitable distribution of a resource across groups in which the distributor observes the effect of the resource only partially. A second approach using MDPs~\citep{wen21} proposes an algorithm in which optimal policies for MDPs are balanced against fairness constraints, using an adaptation of demographic parity to the MDP setting.

\section{Setup}
\label{sec:setup}

\subsection{Timing}
\label{sec:timing}

We will consider time to pass in `epochs' indexed by $e$, within each of which there are two time points $t=0$ and $t=1$. At $t=0$ we observe the covariates of a set of samples. We denote by $X_e(0) \in \mathbb{R}^p$ the covariates of a particular sample, and the set of covariates of all samples $D_e$. We will generally take values $X_e(0)$ to be independent across samples and across $e$ and identically distributed within $e$ (that is, we begin with a `fresh' set of samples at the beginning of each new epoch) but will not be concerned with the distribution of $X_e(0)$.

Immediately after observing $x=X_e(0)$, we may intervene on $x$. We denote the value of $x$ after intervention by $G_e(x)$. At $t=1$, the value of covariates is now $X_e(1)=G_e[X_e(0)]$. The outcome $Y_e$ is determined at $t=1$ with probability depending on $X_e(1)$ through the function $f_e$:
\begin{align}
Y_e &\sim \textrm{Bernoulli}\left\{f_e\left[X_e(1)\right]\right\} \sim \textrm{Bernoulli}\left\{f_e\left[G_e\left(X_e(0)\right)\right]\right\} \nonumber
\end{align}
We then observe the values of $Y_e$ for samples in $D_e$, along with the values $X_e(0)$. 


\subsection{Preliminaries and algorithm}
\label{sec:preliminaries}

We wish to choose our intervention $G_e$ so as to optimise some objective. 
We generally cannot freely modulate $G_e$, as above. Moreover, from only observations $(X_e(0),Y_e)$, the function $G_e$ is generally not identifiable; as a simple example, take $X_e=(X_1,X_2)$, $f(x_1,x_2)=\tanh(x_1 + x_2)$, $G_e(x)=G_e(x_1,x_2)=(\alpha x_1 + (1-\alpha) x_2,(1-\alpha)x_1 + \alpha x_2)$; then 
\begin{equation}
P(Y_e|X_e(0)=(x_1,x_2)) = P(Y_e|X_e(1)=G_e(x_1,x_2)) = f\left(G_(x_1,x_2)\right) =\tanh(x_1 + x_2) \nonumber
\end{equation}
and $\alpha$ (and hence $G_e$) is not identifiable from the joint distribution of $(X_e(0),Y)$ alone. Our approach is essentially to forego identifying $G_e$ while still using it to bring about a population level change. 

We presume that we can recommend a $G_e$ consisting of a composition of consistent responses to a series of risk scores: at $t=0$ we may compute a series of risk scores $\rho_0(x)$, $\rho_1(x)$, $\dots$ and use these risk scores to define 
\begin{equation}
G_e(x)=g\Big[\rho_{e-1}(x),g\big[\rho_{e-2}(x),\dots g[\rho_0(x),x]\dots\big]\Big] \nonumber
\end{equation}
where $g(\rho,x)$ is a function which takes a risk score $\rho$ and a set of covariates $x$ and returns an `intervened-upon' set of covariates. The function $g$ can be thought of as an instruction to expert agents to `act upon $x$ as though the risk of $Y$ given covariates $X$ was $\rho$'. The function $g$ may depend on $X$ only through a subset of values; that is, the agent implementing $g$ may only observe a subset of elements of $X$ when they make the decision to intervene.

Our overall algorithm is recursive. Initially, at epoch $0$, we use no risk score and take $G_e(x)=x$. We then fit a risk score $\rho_0(x)\approx P(Y_0|X_0(0)=x)=f_e(x)$. Thereafter, at epoch $e$, we use risk scores $\rho_0$, $\rho_1$, $\dots$, $\rho_{e-1}$ to define $G_e$. After $t=1$ in epoch $e$ we use observed values $(X_e(0),Y_e)$ to fit a new risk score $\rho_e:\mathbb{R}^p \to [0,1]$, so that
\begin{align}
\rho_e(x) \approx P(Y_e|X_e(0)=x) =P\left[Y_e|X_e(1)=G_e(x)\right] =f_e\left[G_e(x)\right] \nonumber
\end{align}
which we add to our trove of existing scores for use in epoch $e+1$. Often, $f_e$ will not depend on $e$ and we will drop the subscript for convenience. We will not consider variation in $g$ with $e$. Our algorithm is now:

\begin{algorithm}[H]
\KwResult{Intervention $G_E(x)$ on covariates $x$ and predictive score $\rho_{E}(x)$ for use in epoch $E+1$}
\eIf{E=0}{
 $G_{0}(x) \gets x$ \tcp*{Initially no intervention}
 $\rho_0(x) \gets \textrm{estimate of } \mathbb{E}_{Y_0}(Y_0|X_e(0)=x)=f_0(x)$  \tcp*{Fit score}
}{
 $G_E(x) \gets \big\{$
 $\chi_0=x$; 
 \textbf{for} {$e \gets 1$ \textbf{to} $E$ \textbf{do}
  $\chi_{e}=g_E(\rho_{e-1}(x),\chi_{e-1})$; \textbf{return} $\chi_{E}$ $\big\}$
}
 $\rho_e(x) \gets \textrm{estimate of } \mathbb{E}_{Y_e}(Y_e|X_e(0)=x) = f_e\left[G_e(x)\right]$  \tcp*{Refit score}
}
\Return $\rho_{E}$, $g_E$
\caption{Intervention stacking for equitable outcome; epoch $E$}
\label{alg:main}
\end{algorithm}

The instructions to an agent arising to implement $G_e$ are essentially: `Firstly intervene according to $\rho_0$, which is the native risk if no interventions were to be made. Next intervene according to $\rho_1$, which is the estimated risk after intervening on $\rho_0$, then according to $\rho_2$ which is the estimated risk after intervening on $\rho_0$ and $\rho_1$, ... and so on to epoch $E$'. We will show that as we continue to add new risk scores and extend the chain of interventions, the risks of $Y_e$ after the intervention chain will almost converge. 

\subsection{Formulation as POMDP}
\label{sec:pomdp}

When $f_e$ and $g$ do not depend on $e$, and distributions of covariates $X_e(0)$ are identical over $e$, our setting can be described by a Partially Observed Markov Decision Process (POMDP), a common formulation in reinforcement learning~\citep{sutton18}. Following notation from~\citep{wang19}, we define this as a septuple $(S,A,T,R,\Omega,O,\gamma)$, roughly comprising states $S$ in which the agent can be, actions $A$ which it can take, transitions $T$ determining probabilities of transitions between states after a given action, reward $R$ for taking a given action in a given state, observations $\Omega$ governed by observation probability $O$, and a discount factor $\gamma$ for future actions. The generalised aim of a POMDP is to determine a policy $P:\Omega \to A$ which guides actions in such a way as to maximise reward. Our algorithm constitutes such an action policy. The POMDP is specified as: 

\sloppy
\begin{itemize}
\item[$S$]: (states) set of possible covariate values $X$;
\item[$A$]: (actions) set of possible interventions $g(\rho,\cdot)$, for $\rho \in [0,1]$; 
\item[$T$]: (transition prob.) distribution of covariate values after intervention; that is, $T\left(\textrm{end state }x'|\textrm{initial state } x, \textrm{action }g(\rho,\cdot)\right)=P\left(g(\rho,x)=x'\right)$; 
\item[$R$]: (reward) expected variation from an `acceptable' risk level $\rho_{eq}$; $R\left(\textrm{state } x,\textrm{action } g(\rho,\cdot)\right)=\mathbb{E}_{x,G_e}\left\{\left|f[g(\rho,x)] - \rho_{eq}\right|\right\}$
\item[$\Omega$]: (observation) values of $Y_e \in \{0,1\}^n$
\item[$O$]: (observation prob.) probabilities for $Y_e$; e.g. $O\left(\textrm{outcome } Y|\textrm{state } x, \textrm{action }g(\rho,\cdot)\right) = f\left[g(\rho,x)\right]$ 
\item[$\gamma$]: (discount factor) not specified
\end{itemize}
Our policy is to take the series of actions $g(\rho_0,\cdot)$, $g(\rho_1,\cdot)$, $\dots$, with successive $\rho_i$ determined from outcomes $\Omega$. Within a given epoch $e$, we denote the outcome of applying actions $g(\rho_0,\cdot)$, $g(\rho_1,\cdot)$, $\dots$, $g(\rho_i,\cdot)$ to covariates $x$ as $\chi_i$ for $i<e-1$ (as in algorithm~\ref{alg:main}). We will show that this leads to a favourable reward in subsequent sections.

A typical POMDP re-observes observations $\Omega$ after every action. In our setting, in epoch $e$ we do \emph{not} observe outcomes $Y$ after applying $g(\rho_0,\cdot)$, $g(\rho_1,\cdot)$, $\dots$, $g(\rho_i,\cdot)$ for $i<e-1$. Explicitly, at epoch $e$, we observe state $X_e(0)$ before any actions at all, successively apply actions $g(\rho_1,\cdot)$, $g(\rho_2,\cdot)$, \dots, $g(\rho_{e-1},\cdot)$, and only then observe $Y_e$ and determine a new action $g(\rho_e,\cdot)$. 
We then start again with a new observed $X_{e+1}(0)$, on which we then perform the actions $g(\rho_{e-1},g(\rho_{e-2},\dots,g(\rho_1,X_{e+1}(0))\dots))$ and only then perform $g(\rho_e,\cdot)$ on the result. In this sense, we `restart' our series of actions after each new proposed action.

\section{Convergence to equitable outcome}
\label{sec:results}

We will begin this section by stating several regularity assumptions on $f_e$ and $g$. Not all assumptions will be used in every theorem. We begin with an assumption that essentially states that the intervention $g$ has a realistic effect on covariate values:

\begin{assumption}[Closure of $g$]
\label{asm:g_closure}

Denote by $\Xi$ the set of possible values of $X_e(0)$. Then 
there exists a space $\mathscr{X}$ of random variables with domain $\Xi$ such that:
\begin{align}
\delta_{\xi} \in \mathscr{X} \textrm{ for all } \xi \in \Xi \nonumber \\
\mathbb{E}\{\xi\} \textrm{ is finite for all } \xi \in \Xi \nonumber \\
g:\mathscr{X} \to \mathscr{X} \nonumber 
\end{align}
\end{assumption}

In other words, assumption~\ref{asm:g_closure} asserts that function $g$ maps plausible distributions of $X$ (which include singleton values) to other plausible distributions, and preserve existence of the first moment. If $g$ is deterministic, then the assumption reduces to the assertion that $g:\Xi\to\Xi$

Our next assumption specifies an important constraint on the overall behaviour of the intervention. We define this in terms of an `equivocal risk value' $\rho_{eq}$, which can be thought of as an acceptable risk.

\begin{assumption}[Interventions are well-intentioned (at level  $(q,\gamma)$ and with respect to $f$)]
\label{asm:rho_eq}

There exists $\rho_{eq} \in (0,1)$, $\gamma>0$, and a strictly increasing locally Lipschitz function $q(\cdot):(0,1) \to \mathbb{R}$ such that for all $e>0$, $X \in \mathscr{X}$ and $\rho \in [0,1]$:
\begin{alignat}{2}
&\textrm{  If  } \rho \lessgtr \rho_{eq}: \quad
&\pm q\left\{\mathbb{E}\left[f\left(g(\rho,X)\right)\right]\right\}  \mp  q\left\{\mathbb{E}\left[f\left(X\right)\right] \right\} &\geq \gamma|\rho-\rho_{eq}| \nonumber \\  
&\textrm{  If  } \rho = \rho_{eq}: \quad
&\mathbb{E}_{g}\left[f\left(g(\rho,x)\right)\right]  &=  f(x) \ \label{eq:distribution_change}
\end{alignat}
for a function $f:\mathbb{R} \to (0,1)$.
\end{assumption}

If $g$ is deterministic this reduces to the conditions that $g(\rho_{eq},x)=x$ and $ (\rho \lessgtr \rho_{eq}) \Leftrightarrow \pm q\left\{f_e\left(g(\rho,x)\right)\right\} \mp q\left\{f_e(x)\right\} \geq  \gamma|\rho-\rho_{eq}|$ for all $x \in \Xi$.

This assumption essentially states that interventions always act to move covariates in a direction that moves the risk of the outcome towards $\rho_{eq}$ by a non-vanishing amount in expectation. As discussed in section~\ref{sec:introduction}, we argue that this is reasonable for real-world interventions. 
%

Typically we cannot intervene maximally on every sample, and the aim of the predictive risk score is to help distribute interventions to samples at high need. For samples assessed as at higher-than-acceptable risk, we aim to reduce risk to an acceptable level. Action for samples for whom assessed risk is already below the acceptable level is of less importance, and for such samples risk may increase due to interventions being used elsewhere (hence $\mathbb{E}_{X}\{f(X_e)\}$ may increase from $e$ to $e+1$ if $\rho_e<\rho_{eq}$). Should a range of risk values be acceptable, assumptions~\ref{asm:g_closure} can be extended to allow an `equivocal interval' $\rho_{eq}=[\rho_{eq}^0,\rho_{eq}^1]$, and `$\rho=\rho_{eq}$' replaced with `$\rho \in [\rho_{eq}^0,\rho_{eq}^1]$ in subsequent results.

We include the function $q(\cdot)$ to allow for interventions which have small effects on the scale of $\rho$, but have nonetheless non-negligible effects on covariates. An example for which a non-trivial function $q(\cdot)$ is used is shown in supplementary section~\ref{supp_sec:theorem_simulations}

Finally, we specify assumptions that our risk score is an `oracle', in that it exactly estimates the relevant risk, and assume the absence of `drift', in that $f_e$ does not change with $e$. In later sections we will relax these assumptions. 

\begin{assumption}[Oracle $\rho$] 
\label{asm:oracle}

The fitted risk score $\rho_e$ satisfies
%
$\rho_{e}(x) = P(Y_e|X_e(0)=x)$
%
for all $x \in \Xi$.
\end{assumption}

\begin{assumption}[No drift]
\label{asm:no_drift}
Functions $f_e$ and $g$ are the same for all $e$ (and will be referred to as $f$, $g$).
\end{assumption}

\subsection{Deterministic intervention}

We firstly address the case of deterministic $g$. In this case, we show the following result, proved in section~\ref{sec:apx}; that $P(Y_e|X_e(0)=x)$ tends to `almost converge', in that its limit supremum and infimum are bounded, in a general case, and may fully converge if further conditions are met.

\begin{theorem}
\label{thm:deterministic}

Suppose assumptions~\ref{asm:g_closure}, \ref{asm:rho_eq}, \ref{asm:oracle} and \ref{asm:no_drift} hold, and $\rho_e$, $G_e$ evolve as per algorithm~\ref{alg:main}. Suppose further that $g$ is deterministic; that is, $g:[0,1]\times\Xi \to \Xi$. Then for all $x \in \Xi$ we have (noting $\rho_e(x)=P(Y_e|X_e(0)=x)$)
\begin{align}
\inf_{x \in \Xi:f(x)>\rho_{eq}} f\left(g(f(x),x)\right) &\leq \lim \inf_{e} \rho_e(x) \leq \rho_{eq} \leq \lim \sup_e \rho_e(x) \leq \sup_{x \in \Xi:f(x)<\rho_{eq}} f\left(g(f(x),x)\right) \nonumber
\end{align}
and if $\rho_e(x)$ converges, then it converges to $\rho_{eq}$.
\end{theorem}

We note that given assumption~\ref{asm:oracle}, convergence in $\rho_e(x)$ is tantamount to convergence in true risk $P(Y_e|X_e(0)=x)$. The statement of this theorem is thus essentially that: if an intervention is well-intentioned, in that it moves the true risk of an outcome in the correct direction, then repeated application of the intervention interspersed with risk re-estimation will cause risk to (essentially) converge towards the equivocal value or interval. Algorithm~\ref{alg:main} describes how to implement this feedback process through repeated risk modelling.

As well as convergence of risk, we may be interested in convergence of post-intervention covariates; we do not want convergence in risk to come at the price of divergence of covariates. Fixing $X_e(0)=x$ for all $e$, if $X_e(1)$ converges, then so does $\rho_e(x)$. The convergence of $X_e(1)$ as $e \to \infty$ can be determined from properties of $g$ and $f$. 

We note that for deterministic $g$ and under assumptions~\ref{asm:g_closure}, \ref{asm:rho_eq}, \ref{asm:oracle} and \ref{asm:no_drift} that function $G_e(x)$ simply consists of $e-1$ recursive applications of the function $x \to g(f(x),x)$. Hence if there exists a domain $\mathscr{D} \subseteq \Xi$ which is complete and on which $k(x) = g(f(x),x)$ is a contraction with respect to Euclidean distance, then $X_e(1)=G_e(x)$ will converge. Denoting by $x^i$ the $i$th component of $x$ and defining
\begin{equation}
\partial_i(x) = \frac{\partial k}{\partial x^i}(x) = \frac{\partial g}{\partial x^i}(f(x),x) + \frac{\partial f}{\partial x^i}(x)\frac{\partial g}{\partial \rho}(f(x),x)
\end{equation}
assuming these partial derivatives exist, we have if If $\rho_{eq} \in \mathscr{D}$ and $0<\partial_i(x)<1$ for all $x \in \mathscr{D}$, then $k(x)$ will be such a contraction on $\mathscr{D}$.

In realistic circumstances it may not be reasonable to aim to bring all samples to an equivocal risk $\rho_{eq}$. For instance, in predicting the probability of medical emergencies in the general population, the acceptable risk for elderly individuals with known long-term illnesses may be higher than the acceptable risk for young individuals without existing illness. In this case, we may simply divide the population into two or more subcohorts on the basis of covariates, with separate interventions and associated values of $\rho_{eq}$ in each cohort. As long as interventions cannot lead to sample covariates moving out of their cohort, we may apply results in each cohort separately. 

\subsection{Probabilistic intervention}

We firstly extend to the setting where the value of $g$ is a random variable over $\Xi$. We can assure that $\rho_e$ will eventually not be far from $\rho_{eq}$ if $g$ moves a random variable $X$ in the `right' direction in expectation; that is, changes it such that $f(x)$ moves towards $\rho_{eq}$. The following result is proved in section~\ref{sec:apx}.

\begin{theorem}
\label{thm:probabilistic}
Suppose assumptions~\ref{asm:g_closure}, \ref{asm:rho_eq}, \ref{asm:oracle} and \ref{asm:no_drift} hold, and $\rho_e$, $G_e$ evolve as per algorithm~\ref{alg:main}. Then for all $x \in \Xi$ we have (noting $\rho_e(x)=P(Y_e|X_e(0)=x)$)
\begin{align}
\inf_{\begin{smallmatrix} x \in \Xi,\rho \in [0,1] \\ f(x)>\rho_{eq},\rho>\rho_{eq} \end{smallmatrix}} \mathbb{E}_g \left\{f(g(\rho,x))\right\}
\leq \lim \inf \rho_e(x) \leq 
\rho_{eq} \leq \lim \sup \rho_e(x) \leq 
\sup_{\begin{smallmatrix} (x \in \Xi,\rho \in [0,1]) \\ f(x)<\rho_{eq},\rho<\rho_{eq} \end{smallmatrix}} \mathbb{E}_g \left\{f(g(\rho,x))\right\} \nonumber
\end{align}
and if $\rho_e(x)$ converges as $e \to \infty$, then it must converge to $\rho_{eq}$. 
\end{theorem}

We do not want convergence in risk $\rho_e$ to come at the cost of non-convergence in covariate values. We are thus concerned with the behaviour of $X_e(1)$ as $e \to \infty$.

Fixing $X_e(0)$ to a value $x$, if the sequence of random variables $X_e(1)$ converge in distribution then $\rho_e(x)$ converges to $\rho_{eq}$.  The converse (convergence in distribution of $X_e(1)$ given convergence of $\rho_e(x)$) does not hold; an example is shown in Supplementary section~\ref{supp_sec:theorem_simulations}. 

We show that, under a condition guaranteeing convergence of $\rho_{e}(x)$, if the amount of variance added to $X$ by $g(\rho,X)$ decreases sufficiently as $\rho$ gets close to $\rho_{eq}$, then $X_e(1)$ will converge in distribution. Practically, this simply means that interventions have less effect as assessed risk becomes close to acceptable risk. This result is proved in section~\ref{sec:apx}.

\begin{theorem}
\label{thm:convergence}

Suppose that assumptions~\ref{asm:g_closure}, \ref{asm:oracle} and \ref{asm:no_drift} hold, that $\rho_e$ and $G_e$ evolve as per algorithm~\ref{alg:main}, that for all $X \in \mathscr{X}$ we have, denoting $X^1=g(\mathbb{E}_X\{f(X)\},X)$, 
for some strictly increasing locally Lipschitz function $q(\cdot):(0,1) \to \mathbb{R}$:
\begin{equation}
\left|\frac{q\left\{\mathbb{E}\left[f(X^1)\right]\right\}-q(\rho_{eq})}{q\left\{\mathbb{E}\left[f(X)\right]\right\} - q(\rho_{eq})}\right| \leq 1-\epsilon_0 \nonumber
\end{equation}
and that we may decompose $g$ such that for $X \in \mathscr{X}$
\begin{align}
g(\mathbb{E}\{f(X)\},X) \defeq X + Z \nonumber
\end{align}
where $Z$, considered as a function of $\rho=\mathbb{E}\{f(X)\}$, satisfies
\begin{align}
\mathbb{E}(|Z|) &= O_{f,g}\left(|\rho-\rho_{eq}|\right) \nonumber \\
\textrm{var}(|Z|) &= O_{f,g}\left(|\rho-\rho_{eq}|\right) \nonumber
\end{align}
Then $X_e(1)|X_e(0)=x$ converges in distribution, and $\rho_{e}(x)$ converges pointwise to $\rho_{eq}$ as $e \to \infty$.
\end{theorem}

\section{Robustness}
\label{sec:robustness}

So far, we have used assumption~\ref{asm:oracle} in taking $\rho_{e}(x)$ as an oracle (perfect) estimator of $\mathbb{E}_{Y_e}(Y_e|X_e(0) =x)$. We have also assumed that $f(x)=\mathbb{E}_{Y_e}(Y_e|X_e(1) =x)$ remains constant across epochs $e$ as per assumption~\ref{asm:no_drift}. In a setting in which a slowly-changing system is modelled and a large number of samples are available on which to fit $\rho_e$, these assumptions may be reasonable. However, both of these assumptions will fail to exactly hold in many realistic settings: the modelled system will tend to `drift', changing $f$~\citep{davis19,tahmasbi20}, and $f$ will need to be estimated with associated error. There is little reason to update a predictive risk score in a system not affected by drift.

\subsection{Robustness to error in estimation}
\label{sec:error_robustness}

To relax assumption~\ref{asm:oracle} we must differentiate  $\rho_e$ from the quantity it estimates, which we will denote $\rho_e^o$ (`o' for oracle):
\begin{align}
\rho_e^o(x) &\defeq \mathbb{E}_{Y_e}(Y_e|X_e(0) =x) \nonumber \\
&=\mathbb{E}\left(f\left(G_e(x)\right)\right)
\end{align}
As well as randomness intruduced by $g$, we now have an additional source of randomness: the data used in fitting the function $\rho_e$, which we denote $D_e$. The function $\rho_E^o(x)$ depends on the datasets $D_e$ for $e<E$, since risk scores fitted to such $D_e$ govern $G_E$. We denote $D=\cup_{i=0}^{\infty} D_i = \{D_0,D_1,\dots)$.

We are now principally concerned with the behaviour of the sequence $\{\rho_e^o(x),e=1,2,\dots\}$ as $e$ increases. The sequence of values will no longer converge, as new randomness is added at each epoch due to error in $\rho_e$. We show instead that in general the sequence will be `attracted' by an interval around $\rho_{eq}$, in that whenever it leaves the interval, it moves back towards $\rho_{eq}$ in expectation by a non-vanishing amount.


We will require two further assumptions. Firstly, we are dependent on the accuracy of the fitting of $\rho_e$, which we quantify by requiring $\mathbb{E}_g\left\{f(g(\rho,z))\right\}$ to be `nearly'-convex in $\rho$. In general, we cannot impose that $\mathbb{E}_g\left\{f(g(\rho,z))\right\}$ be convex, as in practice $f$ will usually roughly resemble a (non-convex) logistic function.

\begin{assumption}[$\delta$-almost convexity]
\label{asm:delta_almost_convex}

Given $f$, $g$ and a means to fit $\rho_e$ given data $D_e$, we say the system is `$\delta$-almost convex' if for all $x$ we have
\begin{equation}
\left|\mathbb{E}_{g}\left\{f\left(g\left(\mathbb{E}_{D_e}\{\rho_e(x)\},z\right)\right)\right\} - \mathbb{E}_{D_e,g}\left\{f\left(g\left(\rho_e(x),z\right)\right)\right\}\right| \leq \delta \label{eq:delta_almost_convex}
\end{equation}
\end{assumption}

Note that this is also is a bound on the concavity of $\mathbb{E}_{g}\left\{f\left(g(\cdot,z)\right)\right\}$. We note that, in general, for a larger dataset $D_e$, the variance of $\rho_e(x)$ will be smaller, and hence for locally linear $f$, $g$, the smaller $\delta$ can be. This is consistent with the intuition that the model-intervention complex will work better with larger training datasets.

%
%
Finally, although $\rho_e$ can be biased, we require that
it be on the `right' side of $\rho_{eq}$ in expectation by a non-vanishing amount:
\begin{assumption}[$\lambda$-order effect]
\label{asm:lambda_order}

There exists $\lambda>0$ such that
\begin{align}
\frac{\mathbb{E}_{D_e}\left\{\rho_e(x)\right\}-\rho_{eq}}{\rho_e^o(x)-\rho_{eq}} &\geq \lambda 
\label{eq:assumption_rho_expectation_apx}
\end{align}
\end{assumption}

We can now state a version of theorem~\ref{thm:probabilistic} which allows for errors in the estimation of $\rho_e$.  An important note on this result is that errors do not compound across epochs: errors in a previous fitted score are `corrected' in a sense by making further observations. We show that the risk score $\rho_e$ may be substantially biased as an estimator of $\rho_e^o$, while still leading to values $\rho_e^o(x)$ tending to stay near $\rho_{eq}$. The proof (along with a stronger statement that the movement is non-vanishing if $\rho_e^o(x)$ is outside $[I_1,I_2]$ by at least $\epsilon$ is given in Supplement~\ref{sec:apx}.

\begin{theorem}
\label{thm:errors}

Suppose that the system of $f$, $g$, $\rho_e$ satisfies assumptions~\ref{asm:delta_almost_convex} and~\ref{asm:lambda_order} ($\delta$-almost convexity and $\lambda$-order effect) and assumption~\ref{asm:rho_eq} holds at level $\gamma$. Define the interval $[I_1,I_2]=\left[\rho_{eq}- \frac{\delta}{\gamma \lambda},\rho_{eq}+ \frac{\delta}{\gamma \lambda}\right]$. Then the values $\rho_e^o(x)$ form a discrete random process such that
\begin{align}
\textrm{If } \rho_{e-1}^o(x)> I_2 \textrm{ then } \mathbb{E}_{D_e}\{\rho_e^o(x)|D\}<\rho_e^o(x) \nonumber \\
\textrm{If } \rho_{e-1}^o(x)< I_1 \textrm{ then } \mathbb{E}_{D_e}\{\rho_e^o(x)|D\}>\rho_e^o(x) \nonumber 
\end{align}
that is, they `move' towards $[I_1,I_2]$ when they are outside it. 
\end{theorem}

\subsection{Robustness to drift}

We now consider a setting where $f$ may change across epochs $e$. We re-introduce the subscript $e$ to $f$, and firstly consider the situation where each $f_e$ is `close' to an underlying function $f$. 

\begin{assumption}[$(q,\alpha)$-boundedness] We say functions $f_e$ are $\alpha$-bounded if their absolute difference does not exceed $\alpha$ when transformed by a strictly increasing locally Lipschitz function $q:(0,1) \to \mathbb{R}$
\begin{equation}
\forall x,e: |q\{f(x)\} - q\{f_e(x)\}| \leq \alpha \label{eq:fdif}
\end{equation}
\end{assumption}

We presume assumption~\ref{asm:rho_eq} at level $\gamma$ 
with respect to $f$ rather than $f_e$. 
We allow an oracle estimator $\rho_e$ as per assumption~\ref{asm:oracle} and show essentially that theorem~\ref{thm:probabilistic} is moderately robust to drift of this type (proved in section~\ref{sec:apx}): 

\begin{theorem}
\label{thm:drift}
Suppose that $\rho_e$, $G_e$ evolve as per algorithm~\ref{alg:main}, that the functions $f_e$ are $(q,\alpha)$-bounded about some function $f$, that the intervention $g$ is well-intentioned as per assumption~\ref{asm:rho_eq} at level $(q,\gamma)$ with respect to $f$, and assumptions~\ref{asm:g_closure} and \ref{asm:oracle} hold.  Defining 
\begin{align}
I_{\rho} &=\left[\rho_{eq}-2\frac{\alpha}{\gamma},\rho_{eq}+2\frac{\alpha}{\gamma}\right] \nonumber \\
S_{\inf} &= \left\{X \in \mathscr{X}:q\{\mathbb{E}\{f(X)\}\} \leq \alpha + q\left\{\rho_{eq} + \alpha\frac{1+\gamma}{\gamma}\right\}\right\} \nonumber \\
S_{\sup} &= \left\{X \in \mathscr{X}:q\{\mathbb{E}\{f(X)\}\} \geq -\alpha + q\left\{\rho_{eq} - \alpha\frac{1+\gamma}{\gamma}\right\}\right\} \nonumber \\
I_{\lim} &= \left[\inf_{X \in S_{\inf}} q^{-1}\left[-\alpha +  q\left\{\mathbb{E}\left\{f\left(g\left(\mathbb{E}\{f(X)\},X\right)\right)\right\}\right\}\right],  \sup_{X \in S_{\sup}} q^{-1}\left[\alpha +  q\left\{\mathbb{E}\left\{f\left(g\left(\mathbb{E}\{f(X)\},X\right)\right)\right\}\right\}\right]\right] \nonumber
\end{align}
we have that one of the following holds:
\begin{align}
\lim_{e \to \infty} \rho_e(x) &\in \left\{\min(I_{\rho}),\max(I_{\rho})\right\} \nonumber \\
\left[\lim \inf_{e \to \infty} \rho_e(x), \lim \sup_{e \to \infty} \rho_e(x)\right] &\subseteq I_{\lim} \nonumber
\end{align}

\end{theorem}
We note that Theorem~\ref{thm:drift} reduces to Theorem~\ref{thm:probabilistic} when $\alpha=0$. Although this theorem indicates some robustness to deviation of $f_e$ from $f$ (that is, $\alpha>0$) the compromises to Theorem~\ref{thm:probabilistic} are inconvenient, particularly given the appearance of $\lambda$ in the denominator of bounds of $I_{\rho}$, $I_{X}$. In particular, for long-term gradual drift, where $\alpha$ may not be bounded, this theorem is not reassuring on the behaviour of $\rho_e(x)$.

However, the procedure is robust to a form of drift in $f_e$ characterised by intermittent large shifts.  Theorems~\ref{thm:probabilistic}, \ref{thm:errors} and \ref{thm:drift} all demonstrate a `self-correcting' property of algorithm~\ref{alg:main}, in that each new intervention added has a positive effect on $\rho_e(x)$. If we suppose that for $0\leq e<E_1$ we have $f_e=f^1$, and then from $e=E_1$ onwards we have $f_e=f^2\neq f^1$, then conceptually $\rho_e(x)$ will `recover' for $e>E_1$, despite $G_e$ beginning by acting on $E_1$ risk scores fitted to a system with different $f$. This is formalised in the following corollary: essentially if $f_e$ changes via a series of occasional `shocks' at values $E_i$, between which $f_e$ remains reasonably stable, then as long as there are sufficient epochs between these shocks, the values $\rho_e(x)$ will re-stabilise between them.
\begin{corollary}
\label{cor:shocks}

Suppose that $\rho_e$, $G_e$ evolve as per algorithm~\ref{alg:main} and assumptions~\ref{asm:g_closure} and \ref{asm:oracle} hold. Suppose we have some sequence $E_1,E_2,\dots$ where $E_{i+1}-E_i > n_{\min}$ such that for $e \in [E_i,E_{i+1})$ all $f_e$ are $(q,\alpha)$-bounded around a function $f^i$ and that at each epoch $g$ is well-intentioned at level $(q,\gamma)$ with respect to $f^i$. Define $I_{\rho}^i$, $I_{\lim}^i$ as per theorem~\ref{thm:drift} with $f^i$ in place of $f$. Taking some $\epsilon_0>0$, if $n_{\lim}$ is sufficiently large (depending on $\epsilon_0$), there exists some $n_{\epsilon} < n_{\min}$ such that one of
\begin{align}
|\rho_e(x) - \min(I_{\rho}^i)| &< \epsilon_0 \nonumber \\ 
|\rho_e(x) - \max(I_{\rho}^i)| &< \epsilon_0 \nonumber \\ 
\rho_e(x) \in I_{\lim}^i \nonumber
\end{align}
holds for all $e$ with $E_i + n_i \leq e < E_{i+1}$

\end{corollary}

Simulations demonstrating Theorems~\ref{thm:deterministic}, \ref{thm:probabilistic}, \ref{thm:convergence}, \ref{thm:errors}, \ref{thm:drift} and corollary~\ref{cor:shocks} are given in Supplementary section~\ref{supp_sec:theorem_simulations}


\section{Applications}
\label{sec:applications}


\subsection{General simulation}
\label{sec:motivating_example}

Suppose we have the following setting:

\begin{displayquote}
{\bf Scenario.} A population of individuals is assessed in primary care once a year, at which point a range of health data are measured (demographic, lifestyle, medical history). Some will need secondary healthcare at some point during the year, and we want the risk of this to be at an `acceptable' level for all patients. Practitioners can give advice and prescribe medications which have an unknown but `well-intentioned' effect.
\end{displayquote}

We simulated this setting and evaluated the stacked-intervention method as an approach to achieve a good balance of outcomes in the population. Details of the simulation are given in Supplementary section~\ref{supp_sec:example_details}. The approach would implemented by the following instructions to a practitioner: 
\begin{displayquote}
{\bf Action.} At the start of the year you will get a series of scores $\rho_0$, $\rho_1$, $\rho_2$ and so on. Act on $\rho_0$ when you first see the patient. At their next appointent, act on $\rho_1$. The next time, on $\rho_2$, and so on. Each time, treat the risk score $\rho_i$ as the current risk of needing secondary healthcare this year.
\end{displayquote}

\begin{figure}[h]
\begin{subfigure}{0.5\textwidth}
\includegraphics[width=\textwidth]{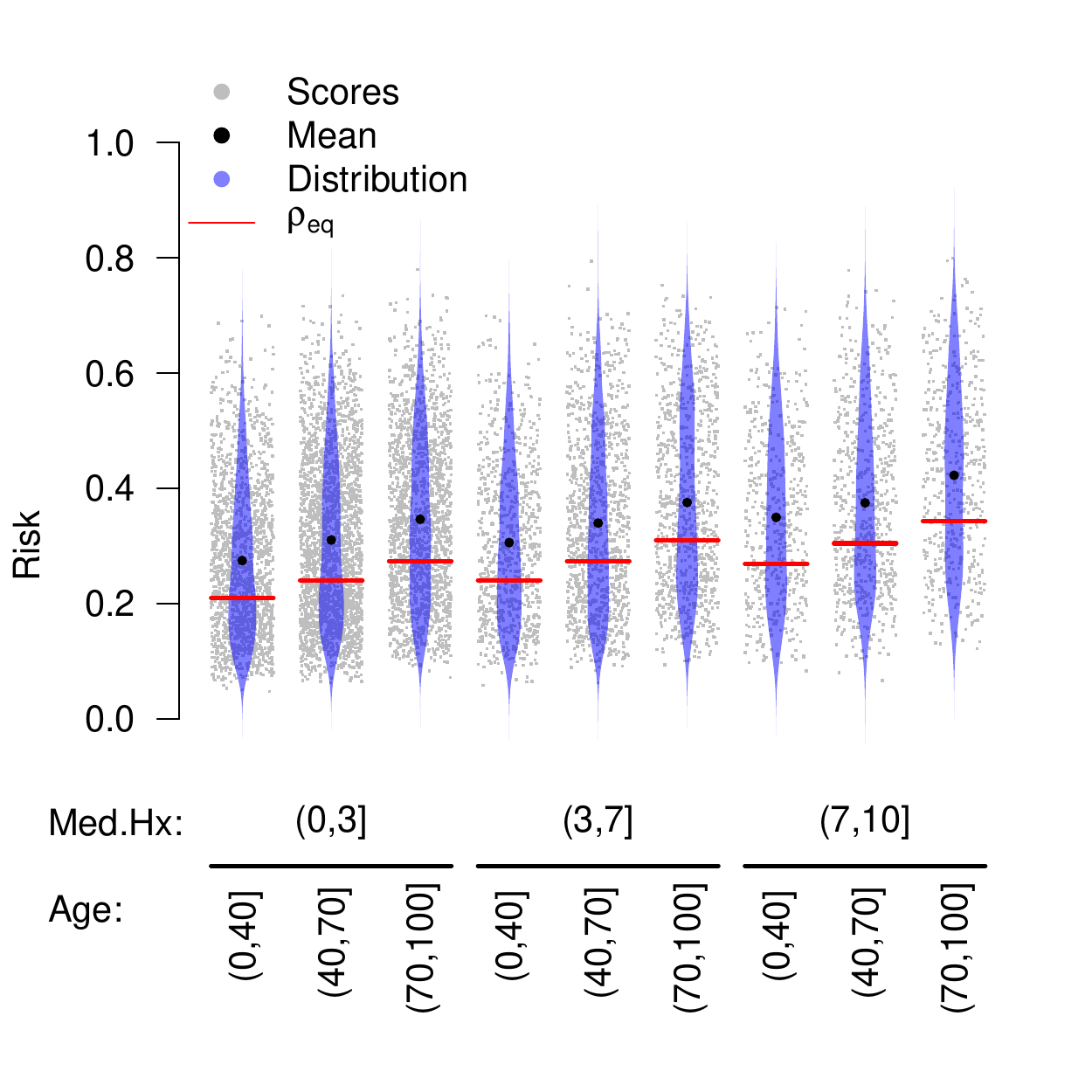}
\end{subfigure}
\begin{subfigure}{0.5\textwidth}
\includegraphics[width=\textwidth]{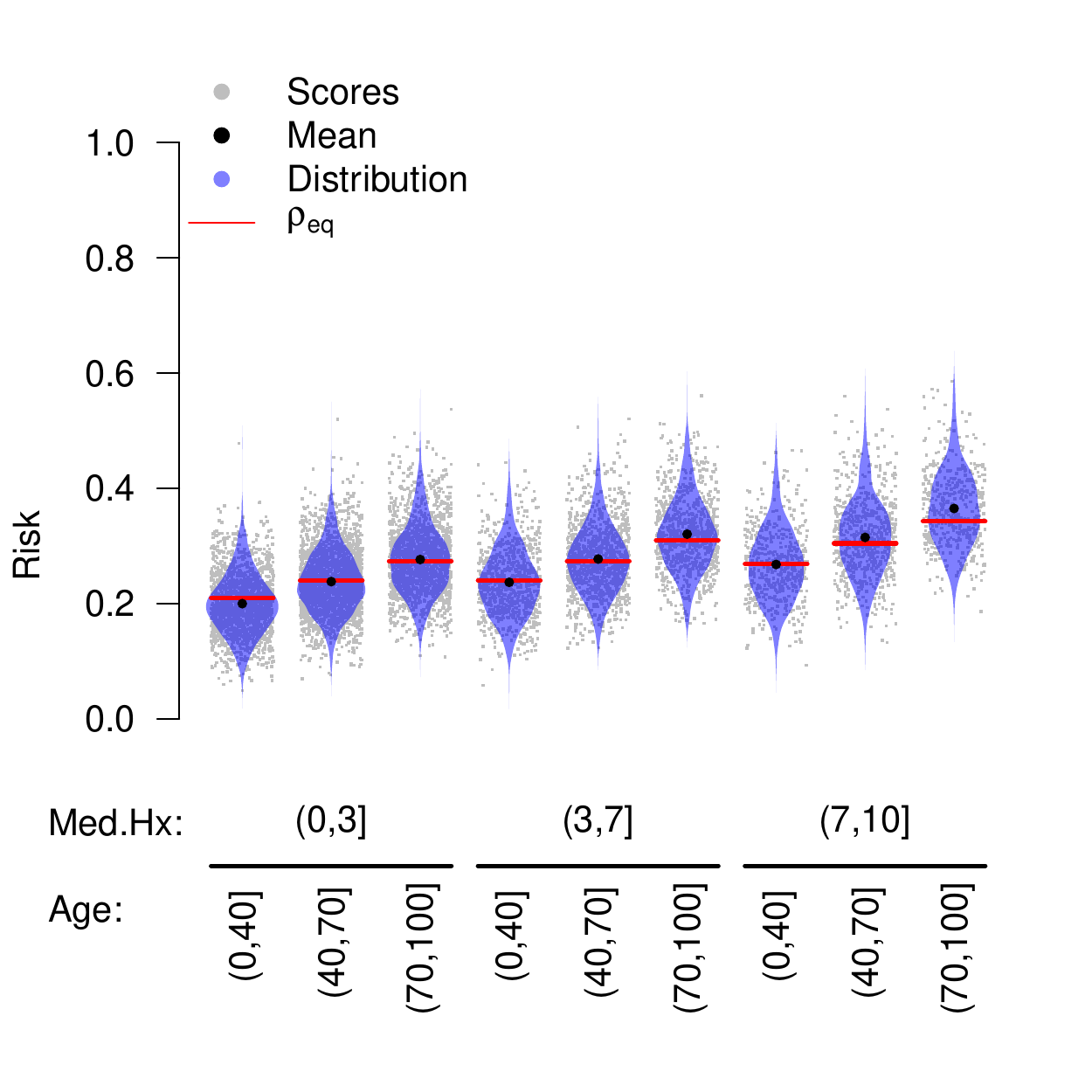}
\end{subfigure}
\caption{Distribution of individual risks before and after intervention under a stacked intervention scheme. Each column is a risk category, defined by age and medical history, with different acceptable risks $\rho_{eq}$ in each. Black dots indicate mean risk in each category.}
\label{fig:main_simulation}
\end{figure}

Figure~\ref{fig:main_simulation} shows the distribution of all true individual risk values before any intervention and after twenty stacked interventions across medical history and age categories. The reduction in variance of risks around equitable values $\rho_{eq}$ is clear. The stacked interventions indicated by the risk score lead to an imperfect convergence of risks towards $\rho_{eq}$.

\section{Potential practical implementations}

\begin{figure}[h]
\centering
\includegraphics[width=0.8\textwidth]{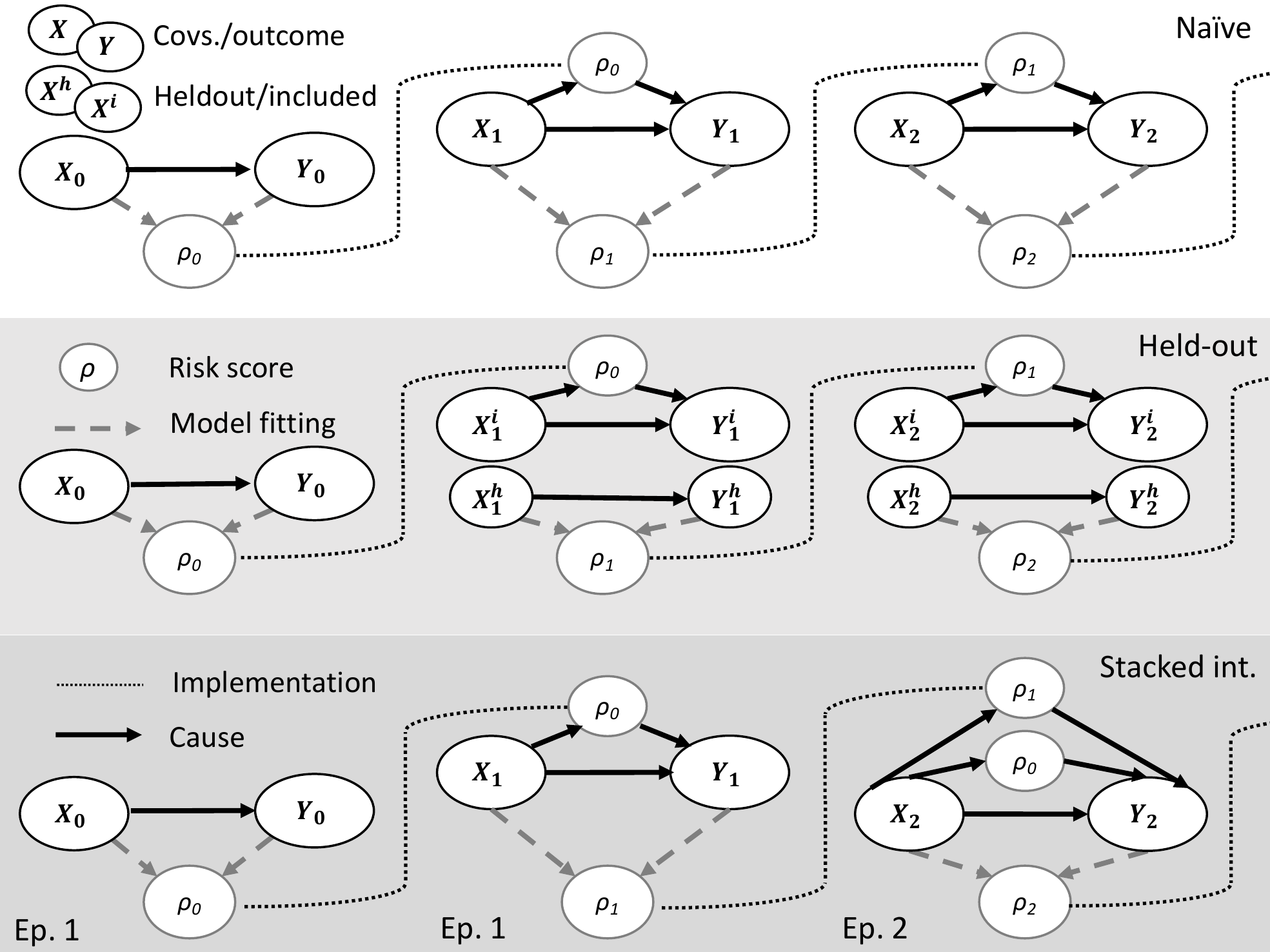}
\caption{Causal graphs associated with naive updating, the use of a hold-out updating strategy, and stacked interventions. Notably for na\"{i}ve updating, scores $\rho_0$, $\rho_1$ are implemented on systems different to those to which they were fitted.}
\label{fig:applications_causal}
\end{figure}

We now discuss two potential applications of stacked interventions by considering how the algorithm may be used in the context of well established predictive risk scores. We contrast our algorithm against two alternative updating strategies: `naive' updating strategy, in which a new score is simply refitted and replaces the old score, ignoring any effects of risk-score driven intervention; and a `holdout' updating strategy, in which a a portion of the population is explicitly exempted from risk-score driven interventions and used to train the next iteration of the model. The three updating strategies are illustrated as causal graphs in figure~\ref{fig:applications_causal}.


The Q-RISK score predicts 10-year risk of heart attack or stroke using non-modifiable risk factors (age, medical history, and ethnicity) and modifiable risk factors (blood pressure, blood lipids, smoking status and medications). It is deployed as a  publically available risk calculator usable by clinicians, and is updated periodically to reflect changing epidemiology~\citep{hippisley17}. It is susceptible to risk-score driven interventions.

The SPARRA series of risk scores predicts one-year risk of emergency hospital admission in the Scottish population on the basis of routinely collected electronic healthcare records~\citep{liley21b}. It is computed directly for (essentially) the entire population and deployed directly to general practitioners for the patients in their care, and is periodically updated.

The stacked intervention algorithm would be applied to each risk score by refitting new risk scores periodically under the current protocol, but releasing a series of risk score calculators (for Q-RISK) or sets of predicted risks (for SPARRA) including both new and old, rather than replacing the old with the new. Practitioners using the score to direct interventions would be instructed to act on the first version of the score at an initial consult, the second version of the score the next time they see the patient, and so on. 

This approach would generally lead to convergence of true outcome risk for each patient towards an `acceptable' level ($\rho_{eq}$) for that patient's category according to non-modifiable risk factors, in a similar way to the simulation in section~\ref{sec:motivating_example}. This approach has direct advantages over other updating strategies: it avoids the need to explicitly evaluate effects of interventions (as would be needed for complete causal modelling) and avoids dangerous bias arising from risk-score driven interventions. A hold-out set approach is not readily implementable for publically-available risk calculators such as QRISK, and for SPARRA would necessitate not calculating scores for a portion of the population. An approach of not updating the risk scores at all would be expected to lead to gradual attenuation of predictive performance over time.

\section{Discussion}


Stacked Interventions is a protocol for deployment of predictive risk scores. It enables calibration of interventions to risk scores, and constitutes a method for safe risk score updating, collectively comprising a risk-score-intervention aggregate which is, in a sense, optimal. It does not require knowledge or inference of intervention effects, and never compels expert agents to act against their judgement. 

The main obstacle to the implementation of our method is the need to distribute multiple risk scores rather than one, and to train expert agents to use them. Nonetheless, it does not require training of agents to interpret risk scores in unusual ways (for instance, in the case of a performatively stable equilibrium, a risk must be interpreted as `the risk after taking a typical action on the basis of itself'): every risk score can be straightforwardly interpreted as the expected risk after a given set of interventions. The method is clearly impractical once the number of risk scores mounts too high, and periodically `restarts' will be necessary, where a new initial risk score $\rho_0$ is trained, potentially through the use of a holdout set approach~\citep{haidar22}.

Our method can be considered most fundamentally as allowing an effector-feedback loop: a deviation in risk from $\rho_{eq}$ is detected using the risk score and effected by the agent, causing a change in risk back toward $\rho_{eq}$. `Stacking' allows the ability to go around the feedback loop repeatedly, rather than just once. In this way, a deviation of risk from $\rho_{eq}$ can be corrected by successive interventions.

Overall, we consider our method to be a straightforwardly implemented resolution to two major problems in deployment of predictive risk scores: safe updating, and calibration of interventions. As predictive scores are increasingly used to guide interventions, these issues will become more important. The question of how to best intervene on the basis of predictive risk scores is critical as \emph{in vitro} models are adapted for clinical use~\citep{topol19}, and our method is useful in this area.

\section{Data and code sharing}

All code to simulate data and generate results in this work is available on \texttt{https://github.com/jamesliley/Stacked\_interventions}

\section{Acknowledgements}

JL was partially supported by Wave 1 of The UKRI Strategic Priorities Fund under the EPSRC Grant EP/T001569/1, particularly the ``Health'' theme within that grant and The Alan Turing Institute, and was partially supported by Health Data Research UK, an initiative funded by UK Research and Innovation, Department of Health and Social Care (England), the devolved administrations, and leading medical research charities. 

\clearpage



\begin{thebibliography}{31}
\providecommand{\natexlab}[1]{#1}
\expandafter\ifx\csname urlstyle\endcsname\relax
  \providecommand{\doi}[1]{doi:\discretionary{}{}{}#1}\else
  \providecommand{\doi}{doi:\discretionary{}{}{}\begingroup
  \urlstyle{rm}\Url}\fi

\bibitem[{Alaa and van~der Schaar(2018)}]{alaa18}
\textsc{Alaa, Ahmed~M} and \textsc{van~der Schaar, Mihaela}.
\newblock Autoprognosis: Automated clinical prognostic modeling via bayesian
  optimization with structured kernel learning.
\newblock \emph{arXiv preprint arXiv:1802.07207} (2018).

\bibitem[{Artzi et~al.(2020)Artzi, Shilo, Hadar, Rossman, Barbash-Hazan,
  Ben-Haroush, Balicer, Feldman, Wiznitzer, and Segal}]{artzi20}
\textsc{Artzi, Nitzan~Shalom}, \textsc{Shilo, Smadar}, \textsc{Hadar, Eran},
  \textsc{Rossman, Hagai}, \textsc{Barbash-Hazan, Shiri}, \textsc{Ben-Haroush,
  Avi}, \textsc{Balicer, Ran~D}, \textsc{Feldman, Becca}, \textsc{Wiznitzer,
  Arnon}, and \textsc{Segal, Eran}.
\newblock Prediction of gestational diabetes based on nationwide electronic
  health records.
\newblock \emph{Nature medicine}, 26(1):71--76 (2020).

\bibitem[{Davis et~al.(2019)Davis, Greevy~Jr, Fonnesbeck, Lasko, Walsh, and
  Matheny}]{davis19}
\textsc{Davis, Sharon~E}, \textsc{Greevy~Jr, Robert~A}, \textsc{Fonnesbeck,
  Christopher}, \textsc{Lasko, Thomas~A}, \textsc{Walsh, Colin~G}, and
  \textsc{Matheny, Michael~E}.
\newblock A nonparametric updating method to correct clinical prediction model
  drift.
\newblock \emph{Journal of the American Medical Informatics Association},
  26(12):1448--1457 (2019).

\bibitem[{Drusvyatskiy and Xiao(2020)}]{drusvyatskiy20}
\textsc{Drusvyatskiy, Dmitriy} and \textsc{Xiao, Lin}.
\newblock Stochastic optimization with decision-dependent distributions.
\newblock \emph{arXiv preprint arXiv:2011.11173} (2020).

\bibitem[{Elzayn et~al.(2019)Elzayn, Jabbari, Jung, Kearns, Neel, Roth, and
  Schutzman}]{elzayn19}
\textsc{Elzayn, Hadi}, \textsc{Jabbari, Shahin}, \textsc{Jung, Christopher},
  \textsc{Kearns, Michael}, \textsc{Neel, Seth}, \textsc{Roth, Aaron}, and
  \textsc{Schutzman, Zachary}.
\newblock Fair algorithms for learning in allocation problems.
\newblock In \emph{Proceedings of the Conference on Fairness, Accountability,
  and Transparency}, pages 170--179 (2019).

\bibitem[{Finlayson et~al.(2020)Finlayson, Subbaswamy, Singh, Bowers, Kupke,
  Zittrain, Kohane, and Saria}]{finlayson20}
\textsc{Finlayson, Samuel~G}, \textsc{Subbaswamy, Adarsh}, \textsc{Singh,
  Karandeep}, \textsc{Bowers, John}, \textsc{Kupke, Annabel}, \textsc{Zittrain,
  Jonathan}, \textsc{Kohane, Isaac~S}, and \textsc{Saria, Suchi}.
\newblock The clinician and dataset shift in artificial intelligence.
\newblock \emph{The New England Journal of Medicine}, pages 283--286 (2020).

\bibitem[{Friedman et~al.(2001)Friedman, Hastie, and Tibshirani}]{friedman01}
\textsc{Friedman, Jerome}, \textsc{Hastie, Trevor}, and \textsc{Tibshirani,
  Robert}.
\newblock \emph{The Elements of Statistical Learning}, volume~1.
\newblock Springer Series in Statistics New York (2001).

\bibitem[{Gottesman et~al.(2018)Gottesman, Johansson, Meier, Dent, Lee,
  Srinivasan, Zhang, Ding, Wihl, Peng et~al.}]{gottesman18}
\textsc{Gottesman, Omer}, \textsc{Johansson, Fredrik}, \textsc{Meier, Joshua},
  \textsc{Dent, Jack}, \textsc{Lee, Donghun}, \textsc{Srinivasan, Srivatsan},
  \textsc{Zhang, Linying}, \textsc{Ding, Yi}, \textsc{Wihl, David},
  \textsc{Peng, Xuefeng}, \textsc{et~al.}
\newblock Evaluating reinforcement learning algorithms in observational health
  settings.
\newblock \emph{arXiv preprint arXiv:1805.12298} (2018).

\bibitem[{Haidar-Wehbe et~al.(2022)Haidar-Wehbe, Emerson, Aslett, and
  Liley}]{haidar22}
\textsc{Haidar-Wehbe, Sami}, \textsc{Emerson, Samuel~R}, \textsc{Aslett,
  Louis~JM}, and \textsc{Liley, James}.
\newblock Optimal sizing of a holdout set for safe predictive model updating.
\newblock \emph{arXiv preprint arXiv:2202.06374} (2022).

\bibitem[{Hippisley-Cox et~al.(2017)Hippisley-Cox, Coupland, and
  Brindle}]{hippisley17}
\textsc{Hippisley-Cox, Julia}, \textsc{Coupland, Carol}, and \textsc{Brindle,
  Peter}.
\newblock Development and validation of qrisk3 risk prediction algorithms to
  estimate future risk of cardiovascular disease: prospective cohort study.
\newblock \emph{bmj}, 357 (2017).

\bibitem[{Hippisley-Cox et~al.(2007)Hippisley-Cox, Coupland, Vinogradova,
  Robson, May, and Brindle}]{hippisley07}
\textsc{Hippisley-Cox, Julia}, \textsc{Coupland, Carol}, \textsc{Vinogradova,
  Yana}, \textsc{Robson, John}, \textsc{May, Margaret}, and \textsc{Brindle,
  Peter}.
\newblock Derivation and validation of qrisk, a new cardiovascular disease risk
  score for the united kingdom: prospective open cohort study.
\newblock \emph{Bmj}, 335(7611):136 (2007).

\bibitem[{Hippisley-Cox et~al.(2008)Hippisley-Cox, Coupland, Vinogradova,
  Robson, Minhas, Sheikh, and Brindle}]{hippisley08}
\textsc{Hippisley-Cox, Julia}, \textsc{Coupland, Carol}, \textsc{Vinogradova,
  Yana}, \textsc{Robson, John}, \textsc{Minhas, Rubin}, \textsc{Sheikh, Aziz},
  and \textsc{Brindle, Peter}.
\newblock Predicting cardiovascular risk in england and wales: prospective
  derivation and validation of qrisk2.
\newblock \emph{Bmj}, 336(7659):1475--1482 (2008).

\bibitem[{Hyland et~al.(2020)Hyland, Faltys, H{\"u}ser, Lyu, Gumbsch, Esteban,
  Bock, Horn, Moor, Rieck et~al.}]{hyland20}
\textsc{Hyland, Stephanie~L}, \textsc{Faltys, Martin}, \textsc{H{\"u}ser,
  Matthias}, \textsc{Lyu, Xinrui}, \textsc{Gumbsch, Thomas}, \textsc{Esteban,
  Crist{\'o}bal}, \textsc{Bock, Christian}, \textsc{Horn, Max}, \textsc{Moor,
  Michael}, \textsc{Rieck, Bastian}, \textsc{et~al.}
\newblock Early prediction of circulatory failure in the intensive care unit
  using machine learning.
\newblock \emph{Nature Medicine}, 26(3):364--373 (2020).

\bibitem[{Izzo et~al.(2021)Izzo, Zou, and Ying}]{izzo21}
\textsc{Izzo, Zachary}, \textsc{Zou, James}, and \textsc{Ying, Lexing}.
\newblock How to learn when data gradually reacts to your model.
\newblock \emph{arXiv preprint arXiv:2112.07042} (2021).

\bibitem[{Komorowski et~al.(2018)Komorowski, Celi, Badawi, Gordon, and
  Faisal}]{komorowski18}
\textsc{Komorowski, Matthieu}, \textsc{Celi, Leo~A}, \textsc{Badawi, Omar},
  \textsc{Gordon, Anthony~C}, and \textsc{Faisal, A~Aldo}.
\newblock The artificial intelligence clinician learns optimal treatment
  strategies for sepsis in intensive care.
\newblock \emph{Nature medicine}, 24(11):1716--1720 (2018).

\bibitem[{Lenert et~al.(2019)Lenert, Matheny, and Walsh}]{lenert19}
\textsc{Lenert, Matthew~C}, \textsc{Matheny, Michael~E}, and \textsc{Walsh,
  Colin~G}.
\newblock Prognostic models will be victims of their own success, unless...
\newblock \emph{Journal of the American Medical Informatics Association},
  26(12):1645--1650 (2019).

\bibitem[{Li and Wai(2021)}]{li21}
\textsc{Li, Qiang} and \textsc{Wai, Hoi-To}.
\newblock State dependent performative prediction with stochastic
  approximation.
\newblock \emph{arXiv preprint arXiv:2110.00800} (2021).

\bibitem[{Liley et~al.(2021{\natexlab{a}})Liley, Bohner, Emerson, Mateen,
  Borland, Carr, Heald, Oduro, Ireland, Moffat et~al.}]{liley21b}
\textsc{Liley, James}, \textsc{Bohner, Gergo}, \textsc{Emerson, Samuel~R},
  \textsc{Mateen, Bilal~A}, \textsc{Borland, Katie}, \textsc{Carr, David},
  \textsc{Heald, Scott}, \textsc{Oduro, Samuel~D}, \textsc{Ireland, Jill},
  \textsc{Moffat, Keith}, \textsc{et~al.}
\newblock Development and assessment of a machine learning tool for predicting
  emergency admission in scotland.
\newblock \emph{medRxiv} (2021{\natexlab{a}}).

\bibitem[{Liley et~al.(2021{\natexlab{b}})Liley, Emerson, Mateen, Vallejos,
  Aslett, and Vollmer}]{liley21a}
\textsc{Liley, James}, \textsc{Emerson, Samuel}, \textsc{Mateen, Bilal},
  \textsc{Vallejos, Catalina}, \textsc{Aslett, Louis}, and \textsc{Vollmer,
  Sebastian}.
\newblock Model updating after interventions paradoxically introduces bias.
\newblock In \emph{International Conference on Artificial Intelligence and
  Statistics}, pages 3916--3924. PMLR (2021{\natexlab{b}}).

\bibitem[{Mendler-D{\"u}nner et~al.(2020)Mendler-D{\"u}nner, Perdomo, Zrnic,
  and Hardt}]{mendler20}
\textsc{Mendler-D{\"u}nner, Celestine}, \textsc{Perdomo, Juan~C},
  \textsc{Zrnic, Tijana}, and \textsc{Hardt, Moritz}.
\newblock Stochastic optimization for performative prediction.
\newblock \emph{arXiv preprint arXiv:2006.06887} (2020).

\bibitem[{Nashef et~al.(2012)Nashef, Roques, Sharples, Nilsson, Smith,
  Goldstone, and Lockowandt}]{nashef12}
\textsc{Nashef, Samer~AM}, \textsc{Roques, Fran{\c{c}}ois}, \textsc{Sharples,
  Linda~D}, \textsc{Nilsson, Johan}, \textsc{Smith, Christopher},
  \textsc{Goldstone, Antony~R}, and \textsc{Lockowandt, Ulf}.
\newblock Euroscore ii.
\newblock \emph{European Journal of Cardio-Thoracic Surgery}, 41(4):734--745
  (2012).

\bibitem[{Perdomo et~al.(2020)Perdomo, Zrnic, Mendler-D{\"u}nner, and
  Hardt}]{perdomo20}
\textsc{Perdomo, Juan}, \textsc{Zrnic, Tijana}, \textsc{Mendler-D{\"u}nner,
  Celestine}, and \textsc{Hardt, Moritz}.
\newblock Performative prediction.
\newblock In \emph{International Conference on Machine Learning}, pages
  7599--7609. PMLR (2020).

\bibitem[{Roques et~al.(1999)Roques, Nashef, Michel, Gauducheau, De~Vincentiis,
  Baudet, Cortina, David, Faichney, Gavrielle et~al.}]{roques99}
\textsc{Roques, F}, \textsc{Nashef, SAM}, \textsc{Michel, P},
  \textsc{Gauducheau, E}, \textsc{De~Vincentiis, C}, \textsc{Baudet, E},
  \textsc{Cortina, J}, \textsc{David, M}, \textsc{Faichney, A},
  \textsc{Gavrielle, F}, \textsc{et~al.}
\newblock Risk factors and outcome in european cardiac surgery: analysis of the
  euroscore multinational database of 19030 patients.
\newblock \emph{European Journal of Cardio-thoracic Surgery}, 15(6):816--823
  (1999).

\bibitem[{Sperrin et~al.(2019)Sperrin, Jenkins, Martin, and Peek}]{sperrin19}
\textsc{Sperrin, Matthew}, \textsc{Jenkins, David}, \textsc{Martin, Glen~P},
  and \textsc{Peek, Niels}.
\newblock Explicit causal reasoning is needed to prevent prognostic models
  being victims of their own success.
\newblock \emph{Journal of the American Medical Informatics Association},
  26(12):1675--1676 (2019).

\bibitem[{Sperrin et~al.(2018)Sperrin, Martin, Pate, Van~Staa, Peek, and
  Buchan}]{sperrin18}
\textsc{Sperrin, Matthew}, \textsc{Martin, Glen~P}, \textsc{Pate, Alexander},
  \textsc{Van~Staa, Tjeerd}, \textsc{Peek, Niels}, and \textsc{Buchan, Iain}.
\newblock Using marginal structural models to adjust for treatment drop-in when
  developing clinical prediction models.
\newblock \emph{Statistics in Medicine}, 37(28):4142--4154 (2018).

\bibitem[{Subbaswamy et~al.(2021)Subbaswamy, Adams, and Saria}]{subbaswamy21}
\textsc{Subbaswamy, Adarsh}, \textsc{Adams, Roy}, and \textsc{Saria, Suchi}.
\newblock Evaluating model robustness and stability to dataset shift.
\newblock In \emph{International Conference on Artificial Intelligence and
  Statistics}, pages 2611--2619. PMLR (2021).

\bibitem[{Sutton and Barto(2018)}]{sutton18}
\textsc{Sutton, Richard~S} and \textsc{Barto, Andrew~G}.
\newblock \emph{Reinforcement Learning, second edition: An Introduction}.
\newblock MIT Press (2018).

\bibitem[{Tahmasbi et~al.(2020)Tahmasbi, Jothimurugesan, Tirthapura, and
  Gibbons}]{tahmasbi20}
\textsc{Tahmasbi, Ashraf}, \textsc{Jothimurugesan, Ellango},
  \textsc{Tirthapura, Srikanta}, and \textsc{Gibbons, Phillip~B}.
\newblock Driftsurf: A risk-competitive learning algorithm under concept drift.
\newblock \emph{arXiv preprint arXiv:2003.06508} (2020).

\bibitem[{Topol(2019)}]{topol19}
\textsc{Topol, Eric~J.}
\newblock High-performance medicine: the convergence of human and artificial
  intelligence.
\newblock \emph{Nature medicine}, 25(1):44--56 (2019).
\newblock ISBN: 1546-170X Publisher: Nature Publishing Group.

\bibitem[{Wang et~al.(2019)Wang, Liu, Wu, Zhu, Du, Fei-Fei, and
  Tenenbaum}]{wang19}
\textsc{Wang, Yunbo}, \textsc{Liu, Bo}, \textsc{Wu, Jiajun}, \textsc{Zhu,
  Yuke}, \textsc{Du, Simon~S}, \textsc{Fei-Fei, Li}, and \textsc{Tenenbaum,
  Joshua~B}.
\newblock {DualSMC}: Tunneling differentiable filtering and planning under
  continuous {POMDPs}.
\newblock \emph{ijcai.org} (2019).

\bibitem[{Wen et~al.(2021)Wen, Bastani, and Topcu}]{wen21}
\textsc{Wen, Min}, \textsc{Bastani, Osbert}, and \textsc{Topcu, Ufuk}.
\newblock Algorithms for fairness in sequential decision making.
\newblock In \emph{International Conference on Artificial Intelligence and
  Statistics}, pages 1144--1152. PMLR (2021).

\end{thebibliography}

\begin{thebibliography}{1}
\providecommand{\natexlab}[1]{#1}
\expandafter\ifx\csname urlstyle\endcsname\relax
  \providecommand{\doi}[1]{doi:\discretionary{}{}{}#1}\else
  \providecommand{\doi}{doi:\discretionary{}{}{}\begingroup
  \urlstyle{rm}\Url}\fi

\bibitem[{McLennan et~al.(2019)McLennan, Noble, Noble, Plunkett, Wright, and
  Gutacker}]{mclennan19}
\textsc{McLennan, David}, \textsc{Noble, Stefan}, \textsc{Noble, Michael},
  \textsc{Plunkett, Emma}, \textsc{Wright, Gemma}, and \textsc{Gutacker, Nils}.
\newblock The {E}nglish indices of deprivation 2019: Technical report (2019).

\end{thebibliography}


\end{bibunit}

\clearpage

\begin{bibunit}[custom]

\setcounter{section}{0}
\setcounter{page}{1}
\renewcommand\thesection{S\arabic{section}}
\renewcommand\thetheorem{S\arabic{theorem}}
\renewcommand\thelemma{S\arabic{lemma}}
\renewcommand\thecorollary{S\arabic{corollary}}

\startcontents[supplement]
\printcontents[supplement]{l}{1}{\section*{Supplementary Materials}\setcounter{tocdepth}{2}}

\clearpage



\section{Proofs}
\label{sec:apx}

\subsection{Theorem~\ref{thm:deterministic} (deterministic formulation)}

\begin{reptheorem}{thm:deterministic}

Suppose assumptions~\ref{asm:g_closure}, \ref{asm:rho_eq}, \ref{asm:oracle} and \ref{asm:no_drift} hold, and $\rho_e$, $G_e$ evolve as per algorithm~\ref{alg:main}. Suppose further that $g$ is deterministic; that is, $g:[0,1]\times\Xi \to \Xi$. Then for all $x \in \Xi$ we have 
(noting $\rho_e(x)=P(Y_e|X_e(0)=x)$)
\begin{align}
\inf_{x \in \Xi:f(x)>\rho_{eq}} f\left(g(f(x),x)\right) &\leq \lim \inf_{e} \rho_e(x) \leq \rho_{eq} \leq \lim \sup_e \rho_e(x) \leq \sup_{x \in \Xi:f(x)<\rho_{eq}} f\left(g(f(x),x)\right) \nonumber
\end{align}
and if $\rho_e(x)$ converges, then it converges to $\rho_{eq}$.
\end{reptheorem}

\begin{proof}
We firstly show that if $\rho_e(x)$ converges, it must converge to $\rho_{eq}$. Given the evolution rules in algorithm~\ref{alg:main} 
\begin{align}
X_{e+1}(1) &= G_{e+1}(x) \nonumber \\
&= g\left(\rho_e(x),X_e(1) \right) \nonumber \\
&= g\left(f(X_e(1)),X_e(1) \right) \nonumber 
\end{align}
noting $\rho_{e}(x)=f(X_{e}(1))$ by assumption~\ref{asm:oracle}.

From the monotonicity of $q$ in assumption~\ref{asm:rho_eq} we have $q(a)>q(b)$ if and only if $a>b$. Thus
\begin{equation}
 (\rho \lessgtr \rho_{eq}) \Leftrightarrow  f_e\left(g(\rho,x)\right) \gtrless  f_e(x)
\end{equation}
For $x$ with $f(x) >\rho_{eq}$ we thus have
\begin{equation}
f\left(g(f(x),x)\right) < f(x) \nonumber
\end{equation}
or $f(x) < \rho_{eq}$ and $f\left(g(f(x),x)\right)>f(x)$ similarly. Thus when $\rho_e(x)>\rho_{eq}$, we have 
\begin{equation}
\rho_{e+1}(x) = f\left(X_{e+1}(1)\right) = f\left(g\left(f(X_e(1)),X_e(1) \right)\right)< f(X_e(1)) = \rho_{e}(x) \label{eq:rho_direction}
\end{equation}
and similarly for $\rho_e(x)<\rho_{eq}$ we have $\rho_{e+1}(x) > \rho_e(x)$. 

Suppose $\rho_e(x)$ converged to $\rho_{lim}>\rho_{eq}$. Then since $q$ in assumption~\ref{asm:rho_eq} is Lipschitz, $q(\rho_e(x))$ must also converge to $q(\rho_{eq})$. From inequality~\ref{eq:rho_direction}, $\rho_e(x)$ and hence $q\{\rho_e(x)\}$ must be decreasing. Thus for any $\epsilon>0$ we may choose sufficiently large $E$ such that for $e>E$ we have
\begin{equation}
 q\{\rho_{e-1}(x)\}-q\{\rho_e(x)\} < \epsilon(\rho_{lim}-\rho_{eq}) \nonumber 
\end{equation}
but from assumption~\ref{asm:rho_eq}:
\begin{align}
 q\{\rho_{e-1}(x)\}-q\{\rho_e(x)\} &= q\left\{f(g(\rho_e(x),x))\right\} - q\left\{f(x)\right\} \nonumber \\
 &\geq \gamma(\rho_e(x)-\rho_{eq}) \nonumber \\
 &\geq \gamma(\rho_{lim} - \rho_{eq}) \nonumber
\end{align}
which is a contradiction for $\epsilon<\gamma$, with a similar argument for $\rho_{lim}<\rho_{eq}$. Thus if $\rho_e(x)$ converges, it converges to $\rho_{eq}$.

The sequence $\rho_e(x) = f(X_e(1))=P(Y_e|X_e(0)=x)$ either converges or must increase and decrease infinitely often. Thus $\rho_e(x)$ either converges to $\rho_{eq}$ or must be on either side of $\rho_{eq}$ infinitely often (hence $\lim\inf \rho_e(x) \leq \rho_{eq} \leq \lim \sup \rho_e(x)$). The furthest $f(X_e(1))$ can be from $\rho_{eq}$ is immediately after it has `crossed' $\rho_{eq}$, since it must move towards $\rho_{eq}$ immediately afterward by assumption~\ref{asm:rho_eq}. Thus the interval $[\lim \inf f(X_e(1)), \lim \sup f(X_e(1))]$ is limited in width by how much $f\left(g(\rho,x)\right)$ can increase/decrease $x$ when $\rho$ is less than/greater than $\rho_{eq}$ respectively and we have
\begin{align}
\lim\inf P(Y_e|X_e(0)=x) = \lim \inf \rho_e(x) &\geq \inf_{e:f(X_e(1))>\rho_{eq}} f\left(g(\rho_e,X_e(1))\right) \nonumber \\
&\geq \inf_{x \in S:f(x)>\rho_{eq}} f\left(g(f(x),x)\right) 
\end{align}
with a similar argument for $\lim \sup P(Y_e|X_e(0)=x)$. If $\rho_E=\rho_{eq}$ for some $E$, then $\rho_e=\rho_{eq}$ for all $e \geq E$ by assumption~\ref{asm:rho_eq}. 

\end{proof}

\clearpage

\subsection{Theorem~\ref{thm:probabilistic} (probabilistic formulation)}

\begin{reptheorem}{thm:probabilistic}
Suppose assumptions~\ref{asm:g_closure}, \ref{asm:rho_eq}, \ref{asm:oracle} and \ref{asm:no_drift} hold, and $\rho_e$, $G_e$ evolve as per algorithm~\ref{alg:main}. Then for all $x \in \Xi$ we have (noting $\rho_e(x)=P(Y_e|X_e(0)=x)$)
\begin{align}
\inf_{\begin{smallmatrix} x \in \Xi,\rho \in [0,1] \\ f(x)>\rho_{eq},\rho>\rho_{eq} \end{smallmatrix}} \mathbb{E}_g \left\{f(g(\rho,x))\right\} \leq \lim \inf \rho_e(x) \leq 
\rho_{eq} \leq \lim \sup \rho_e(x) \leq 
\sup_{\begin{smallmatrix} (x \in \Xi,\rho \in [0,1]) \\ f(x)<\rho_{eq},\rho<\rho_{eq} \end{smallmatrix}} \mathbb{E}_g \left\{f(g(\rho,x))\right\} \nonumber
\end{align}
and if $\rho_e(x)$ converges as $e \to \infty$, then it must converge to $\rho_{eq}$. 
\end{reptheorem}

\begin{proof}

Again we begin by proving that if $P(Y_e|X_e(0)=x)$ converges as $e \to \infty$, then it must converge to $\rho_{eq}$.  Suppose $X_e(0)=x$ for all $e$. We have
\begin{align}
X_{e+1}(1) &= G_{e+1}(x) \nonumber \\
&= g(\rho_e (x),X_e(1)) \nonumber \\
&= g\left(\mathbb{E}\left\{f\left(X_e(1)\right)\right\},X_e(1) \right) \nonumber
\end{align}
We have $\rho_e(x) = \mathbb{E}\{f(X_e(1))\}$, and hence if $\rho_e(x)>\rho_{eq}$, then from the monotonicity of $q$ in assumption~\ref{asm:rho_eq} we have:
\begin{alignat}{2}
&\phantom{\Leftrightarrow}\quad &q\left\{\mathbb{E}[f(X)]\right\} &\geq q\left\{\mathbb{E}\left[f\left(g(\rho,x)\right)\right]\right\} \nonumber \\ 
&\Rightarrow\quad & \mathbb{E}_{X_e(1)}\{f(X_e(1))\} &\geq \mathbb{E}_{g,X_e(1)}\left[f\left(g(\rho_e(x),X_e(1))\right)\right] \nonumber \\ 
&\Rightarrow\quad & \rho_e(x) &\geq \rho_{e+1}(x) \nonumber
\end{alignat}
and likewise if $\rho_e(x)<\rho_{eq}$ then $\rho_{e+1}(x)> \rho_{e}$.

Similarly to the proof of theorem~\ref{thm:deterministic}, if $\rho_e(x) \to \rho_{lim} > \rho_{eq}$ then $q(\rho_e) \to q(\rho_{lim})>q(\rho_{eq})$. So given $\epsilon>0$ we may choose exists $E$ such that for all $e>E$:
\begin{equation}
q\{\rho_{e-1}(x)\} - q\{\rho_e(x)\} < \epsilon(\rho_{lim}-\rho_{eq})
\end{equation}
but from assumption~\ref{asm:rho_eq}:
\begin{align}
q\{\rho_{e}(x)\} - q\{\rho_{e+1}(x)\} &= q\left\{\mathbb{E}[f(X_e(1)]\right\}-q\left\{\mathbb{E}[g(\rho_e(x),X_e(1))]\right\} \nonumber \\
&\geq \gamma|\rho_e(x)-\rho_{eq}| \nonumber \\
&\geq \gamma|\rho_{lim} - \rho_{eq}| \nonumber
\end{align}
giving a contradiction for $\epsilon<\gamma$. Likewise, $\rho_e(x)$ cannot converge to $\rho_{lim}<\rho_{eq}$.

Similarly to the proof of theorem~\ref{thm:deterministic},  $\rho_e(x)$ is always `moved' towards $\rho_{eq}$, so we must have $\lim \inf \rho_e(x) \leq \rho_{eq} \leq \lim\sup \rho_e(x)$. The movement towards $\rho_{eq}$ may `overshoot' in that if $\rho_e(x)>\rho_{eq}$, we may have $\rho_{e+1}(x)<\rho_{eq}$. The minimum value $\rho_{e+1}(x)$ can take after such overshooting is
\begin{align}
\inf_{e:\mathbb{E}\{f(X_e(1))\}>\rho_{eq}}  \mathbb{E} \left\{f(X_{e+1}(1))\right\} &= \inf_{e:\mathbb{E}\{f(X_e(1))\}>\rho_{eq}}  \mathbb{E}_{X_e(1),g} \left\{f(g(\mathbb{E}\{f(X_e(1))\},X_e(1)))\right\} \nonumber \\
&\geq \inf_{X \in \mathscr{X}:\mathbb{E}\{f(X)\}>\rho_{eq}}  \mathbb{E}_{X,g} \left\{f(g(\mathbb{E}\{f(X)\},X))\right\} \nonumber \\
&\geq \inf_{\begin{smallmatrix} x \in \Xi,\rho \in [0,1] \\ f(x)>\rho_{eq},\rho>\rho_{eq} \end{smallmatrix}} \mathbb{E}_g \left\{f(g(\rho,x))\right\} \nonumber
\end{align}
with an analogous result for $\rho_e(x)<\rho_{eq}$. These values must bound $\rho_e(x)$ for $e$ sufficiently large $e$ which is slightly stronger than the necessary result.

\end{proof}

\clearpage

\subsection{Theorem~\ref{thm:convergence} (convergence of covariate distribution)}

\begin{reptheorem}{thm:convergence}

Suppose that assumptions~\ref{asm:g_closure}, \ref{asm:rho_eq}, \ref{asm:oracle} and \ref{asm:no_drift} hold, that $\rho_e$ and $G_e$ evolve as per algorithm~\ref{alg:main}, that for all $X \in \mathscr{X}$ we have, denoting $X^1=g(\mathbb{E}_X\{f(X)\},X)$,  
for some strictly increasing locally Lipschitz function $q(\cdot):(0,1) \to \mathbb{R}$:
\begin{equation}
\left|\frac{q\left\{\mathbb{E}\left[f(X^1)\right]\right\}-q(\rho_{eq})}{q\left\{\mathbb{E}\left[f(X)\right]\right\} - q(\rho_{eq})}\right| \leq 1-\epsilon_0 \label{eq:hyp_cond1}
\end{equation}
and that we may decompose $g$ such that for $X \in \mathscr{X}$
\begin{align}
g(\mathbb{E}\{f(X)\},X) \defeq X + Z \label{eq:hyp_cond2}
\end{align}
where $Z$, considered as a function of $\rho=\mathbb{E}\{f(X)\}$, satisfies
\begin{align}
\mathbb{E}(|Z|) &= O_{f,g}\left(|\rho-\rho_{eq}|\right) \label{eq:zcond} \\
\textrm{var}(|Z|) &= O_{f,g}\left(|\rho-\rho_{eq}|\right) \label{eq:zvarcond}
\end{align}
%
Then $X_e(1)|X_e(0)=x$ converges in distribution, and $\rho_{e}$ converges pointwise to $\rho_{eq}$ as $e \to \infty$.
\end{reptheorem}

\begin{proof}
We will denote $X_e(0)$ by $x$ throughout. If $X=X_e(1)$ then 
\begin{align}
X^1 &= g(\mathbb{E}_{X_e(1)}\{f(X_e(1))\},X_e(1)) \nonumber \\
&= g(\rho_e(x),X_e(1)) \nonumber \\
&= X_{e+1}(1) \nonumber
\end{align}
so
\begin{alignat}{2}
&\phantom{\Leftrightarrow}\quad &\left|q\left\{\mathbb{E}_{g,X}\left[f(X^1)\right]\right\}-q(\rho_{eq})\right|
&\leq (1-\epsilon_0)\left|q\left\{\mathbb{E}_X\left[f(X)\right]\right\} -q(\rho_{eq})\right|\nonumber \\
&\Rightarrow\quad &\left|q\left\{\mathbb{E}_{g}\left[f\left(X_{e+1}(1)\right)\right]\right\} - q(\rho_{eq})\right|
&\leq  \left(1-\epsilon_0 \right) \left|q\left\{\mathbb{E}_g\left[f(X_e(1))\right]\right\}-q(\rho_{eq})\right| \nonumber \\
&\Rightarrow\quad 
&|q(\rho_{e+1}) - q(\rho_{eq})|
&\leq  \left(1-\epsilon_0 \right) \left|q(\rho_e)-q(\rho_{eq})\right| \label{eq:rho_lt_rhoeq_pos}
\end{alignat}
and
\begin{equation}
|q(\rho_{e})-q(\rho_{eq})| \leq (1-\epsilon_0)|q(\rho_{e-1})-q(\rho_{eq})| \leq (1-\epsilon_0)^e|q(\rho_{0})-q(\rho_{eq})| = O\left((1-\epsilon_0)^e\right)
\end{equation}
and thus $q(\rho_e)$ converges to $q(\rho_{eq})$ at $O_{f,g}\left((1-\epsilon_0)^e\right)$. Given that $q(\cdot)$ is locally Lipschitz and strictly increasing, we have $\rho_e$ converges to $\rho_{eq}$.

Denote the characteristic function of $X$ by $\varphi_{X}(t)$. We will show that, as $e \to \infty$,
\begin{align}
\varphi_{X_e(1)}(t) \to \varphi(t)
\end{align}
pointwise, and show that $\textrm{var}\{|X_e(1)|\}$ is bounded so $X_e(1)$ is tight, from which the result follows by Levy's theorem on characteristic functions. 

From conditions~\ref{eq:zcond}, \ref{eq:zvarcond} and Chebyshev's inequality, for some constant $C_{f,g}$ depending on $f,g$ (but not $e$), given $\epsilon>0$ we may choose $\delta$ with
\begin{equation}
\delta \leq \mathbb{E}(Z) + \sqrt{\frac{1}{\epsilon}}\textrm{var}(Z) \leq \left(1+\sqrt{\frac{1}{\epsilon}}\right)C_{f,g}(1-\epsilon_0)^e \label{eq:deltabound}
\end{equation}
%
such that that for sufficiently large $e$ we have 
\begin{align}
P(|Z| \leq \delta)\geq 1-\epsilon
\end{align}
Now
\begin{align}
\Delta_e &\defeq |\varphi_{X_{e+1}(1)}(t) - \varphi_{X_e(1)}(t)| \nonumber \\
&= |\varphi_{g\left(\mathbb{E}\left\{f(X_e(1))\right\},X_e(1)\right)}(t) - \varphi_{X_e(1)}(t)| \nonumber \\
&= |\varphi_{X_{e}(1)+Z}(t) - \varphi_{X_e(1)}(t)| \nonumber \\
&= \left|\mathbb{E}\left\{e^{i t^T (X_e(1)+Z)}\right\} - \mathbb{E}\left\{e^{i t^T X_e(1)}\right\}\right| \nonumber \\
&= \left|\mathbb{E}\left\{e^{i t^T X_e(1)}\left(e^{i t^T Z}-1\right)\right\}\right| \nonumber \\
&=\left|\int e^{i t^T x}\left(e^{i t^T z}-1\right) df_{X_e(1),Z}\right| \nonumber \\
&= \left|\int_{|z| \leq \delta} \int_{\mathbb{R}^p} e^{i t^T x}\left(e^{i t^T z}-1\right) f_{X_e(1),Z}(x,z) dx dz \right. \nonumber \\
&\phantom{=} \left. + \int_{|z| > \delta} \int_{\mathbb{R}^p} e^{i t^T x}\left(e^{i t^T z}-1\right) f_{X_e(1),Z}(x,z) dx dz\right| \nonumber \\
&=|T_1 + T_2| \leq |T_1| + |T_2| \label{eq:zbound}
\end{align}
Firstly bounding $|T_2|$:
\begin{align}
|T_2|&=\left|\int_{|z| > \delta} \int_{\mathbb{R}^p} e^{i t^T x}\left(e^{i t^T z}-1\right) f_{X_e(1)|Z=z}(x) dx \, f_Z(z) dz\right| \nonumber \\
&\leq\left|\int_{|z| > \delta} \int_{\mathbb{R}^p} 2 f_{X_e(1)|Z=z}(x) dx \, f_Z(z) dz\right| \nonumber \\
&= \left|\int_{|z| > \delta} 2 f_Z(z) dz\right| \nonumber \\
&\leq 2 \epsilon \label{eq:bound2}
\end{align}
by~\ref{eq:deltabound}, and bounding $T_1$ for $|t|< \frac{\pi}{\delta}$ 
\begin{align}
|T_1|&=\left|\int_{|z| \leq \delta} \int_{\mathbb{R}^p} e^{i t^T x}\left(e^{i t^T z}-1\right) f_{X_e(1),Z}(x,z) dx dz \right| \nonumber \\
&\leq \left|\int_{|z| \leq \delta} \int_{\mathbb{R}^p} \left| e^{i t^T x}\left(e^{i t^T z}-1\right) \right| f_{X_e(1),Z}(x,z) dx dz \right| \nonumber \\
&\leq \left|\int_{|z| \leq \delta} \int_{\mathbb{R}^p} \left| 1\left(e^{i t^T \delta}-1\right)\right| f_{X_e(1),Z}(x,z) dx dz \right| \nonumber \\
&= \left|\int_{|z| \leq \delta} \int_{\mathbb{R}^p} \sqrt{2}\sqrt{1-\cos(\delta |t|)} f_{X_e(1),Z}(x,z) dx dz \right| \nonumber \\
&\leq \left|\int_{|z| \leq \delta} \delta |t| f_{Z}(z) dz \right| \nonumber \\
&\leq \delta |t|
\end{align}
where we have made use of the observations that if $|z| \leq \delta$, then $t^Tz\leq \delta |t|$; if $0\leq|t_1|\leq|t_2|\leq\pi$ then $|e^{i t_1}-1|\leq |e^{i t_2}-1|$; and that $1-\cos(x) \leq \frac{1}{2}x^2$.

Now choosing $\epsilon=\epsilon_1^e$ for some value $\epsilon_1$ with $1-\epsilon_0 < \sqrt{\epsilon_1} < 1$, we have
\begin{align}
\Delta_e &\leq |T_1|+|T_2| \nonumber \\
&\leq 2 \epsilon + \delta |t| \nonumber \\
&\leq 2 \epsilon +  \left(1+\sqrt{\frac{1}{\epsilon}}\right)C_{f,g}\left((1-\epsilon_0)^e\right)|t| \nonumber \\
&= 2 \epsilon_1^e +  C_{f,g}\left((1-\epsilon_0)^e + \left(\frac{1-\epsilon_0}{\sqrt{\epsilon_1}}\right)\right)^e|t| \nonumber \\
&= 2 \epsilon_1^e +  C_{f,g}(\epsilon_2^e + \epsilon_3^e)|t| \nonumber \\
\end{align}
where $\epsilon_2,\epsilon_3<1$. Hence $\Delta_e$ decreases geometrically, and $\varphi_{X_e(1)}(t)$ converges pointwise.

Now 
\begin{align}
\mathbb{E}\{|X_{e+1}(1)|\} - \mathbb{E}\{|X_e(1)|\} &= \mathbb{E}\left\{\left|g\left(\mathbb{E}\left\{f(X_e(1))\right\},X_e(1)\right)\right|\right\} - \mathbb{E}\{|X_e(1)|\} \nonumber \\
&= \mathbb{E}\left\{|X_e(1) + Z|\right\} - \mathbb{E}\{X_e(1)\} \nonumber \\
&\leq \mathbb{E}\{|X_e(1)|\} + \mathbb{E}\left\{|Z|\right\} - \mathbb{E}\{|X_e(1)|\} \nonumber \\
&\leq \mathbb{E}\{|Z|\} \nonumber \\
&= O\left(\left|\rho_e-\rho_{eq}\right|\right) \nonumber \\
&= O\left((1-\epsilon_0)^e\right)
\end{align}
so $\mathbb{E}\{|X_e(1)|\}$ converges to a finite value, and
\begin{align}
\textrm{var}\{|X_{e+1}(1)|\}  &= \textrm{var}\{|X_e(1) + Z|\} \nonumber \\
&\leq \textrm{var}(|X_e(1)|) + \textrm{var}(|Z|) + 2 |\mathbb{E}(Z^T X_e(1))| + 2\mathbb{E}(|Z|)\mathbb{E}(|X_e(1)|) \nonumber \\
&\leq \textrm{var}(|X_e(1)|) + 2 |\mathbb{E}(Z^T X_e(1))| + O\left((1-\epsilon_0)^e\right) \nonumber \\
&\leq \textrm{var}(|X_e(1)|) + 2 \mathbb{E}(|Z^T| |X_e(1)|) + 2\mathbb{E}(|Z|)\mathbb{E}(|X_e(1)|) + O\left((1-\epsilon_0)^e\right) \nonumber \\
&\leq \textrm{var}(|X_e(1)|) + 2\textrm{cov}(|Z|,|X_e(1)|) + 4\mathbb{E}(|Z|)\mathbb{E}(|X_e(1)|) + O\left((1-\epsilon_0)^e\right) \nonumber \\
&\leq \textrm{var}(|X_e(1)|) + 2\textrm{var}(|Z|)\textrm{var}(|X_e(1)|) + 2\mathbb{E}(|Z|)\mathbb{E}(|X_e(1)|) + O\left((1-\epsilon_0)^e\right) \nonumber \\
&= \textrm{var}(|X_e(1)|)(1+O\left((1-\epsilon_0)^e\right)) + O\left((1-\epsilon_0)^e\right) 
\end{align}
so $\textrm{var}(|X_e(1)|)$ converges also. Thus $X_e(1)$ is tight, which completes the proof.
\end{proof}

\clearpage

\subsection{Theorem~\ref{thm:errors} (error in estimation)}

\begin{reptheorem}{thm:errors}

Suppose that the system of $f$, $g$, $\rho_e$ satisfies assumptions~\ref{asm:delta_almost_convex} and~\ref{asm:lambda_order} ($\delta$-almost convexity and $\lambda$-order effect) and assumption~\ref{asm:rho_eq} holds at level $\gamma$. Define the interval $[I_1,I_2]=\left[\rho_{eq}- \frac{\delta}{\gamma \lambda},\rho_{eq}+ \frac{\delta}{\gamma \lambda}\right]$. Then the values $\rho_e^o(x)$ form a discrete random process such that
\begin{align}
\textrm{If } \rho_{e-1}^o(x)> I_2 \textrm{ then } \mathbb{E}_{D_e}\{\rho_e^o(x)|D\}<\rho_e^o(x) \nonumber \\
\textrm{If } \rho_{e-1}^o(x)< I_1 \textrm{ then } \mathbb{E}_{D_e}\{\rho_e^o(x)|D\}>\rho_e^o(x) \nonumber 
\end{align}
that is, they `move' towards $[I_1,I_2]$ when they are outside it. Moreover, if values $\rho_e^o(x)$ are outside $[I_1,I_2]$ by at least $\epsilon>0$, then the values move by a non-vanishing amount; we have
\begin{equation}
\mathbb{E}_{D_e}\{\rho_e^o(x)|\cup_{i=0}^{e-1} D_{i}\} \in \begin{cases} 
[J_1,\rho_{e}^o(x) - \gamma \lambda \epsilon] &\textrm{ if } \rho_{e-1}^o(x)>I_2 + \epsilon \\
[\rho_{e-1}^o(x) + \gamma \lambda \epsilon, J_2] &\textrm{ if } \rho_{e-1}^o(x)<I_1 - \epsilon \\
[J_1,J_2] &\textrm{ if } \rho_e^o(x)\in [I_1,I_2] 
\end{cases}
\label{eq:thm_errors_main}
\end{equation}
where 
\begin{align}
J_1 &= -\delta + \inf_{\begin{smallmatrix} \rho \geq \rho_{eq} + \lambda(I_1-\rho_{eq}) \\ X:\mathbb{E}\{f(X)\} \geq I_1 \end{smallmatrix}} \mathbb{E}_{g}\left\{f\left(g(\rho,X)\right)\right\} \nonumber \\
J_2 &= \delta + \sup_{\begin{smallmatrix} \rho \leq \rho_{eq} + c_1(I_2-\rho_{eq}) \\ X:\mathbb{E}\{f(X)\} \leq I_2 \end{smallmatrix}} \mathbb{E}_{g}\left\{f\left(g(\rho,X)\right)\right\} \nonumber
\end{align}

\end{reptheorem}

\begin{proof}

Since in statement~\ref{eq:thm_errors_main} we are conditioning on $D_0$, $D_1$, $\dots$, $D_{e-1}$, we can consider $\rho_0$, $\rho_1$, $\dots$, $\rho_{e-1}$ fixed rather than random. As in previous proofs, we will consider the sequence of values $X_1(1)$, $X_2(1)$, $\dots$ given that $X_e(0)=x$ for all $e$. 
%
%
We have
\begin{equation}
P(Y_e|X_e(0)=x) = \mathbb{E}_g\left\{f(X_e(1))\right\} = \rho_e^o(x)
\end{equation}

Suppose $\rho^o_e(x) \geq \rho_{eq} + \frac{\delta}{\gamma \lambda}$. Then from assumption~\ref{eq:assumption_rho_expectation_apx} we have 
\begin{equation}
\mathbb{E}_{D_e}\left\{\rho_e(x)\right\}  - \rho_{eq} \geq \lambda(\rho_e^o(x)-\rho_{eq}) \geq \frac{\delta}{\gamma} \label{eq:rho_e_lower}
\end{equation}
Now
\begin{align}
\mathbb{E}_{D_e}\left\{\rho_{e+1}^o(x)\right\} &= \mathbb{E}_{D_e,g}\left\{f(X_{e+1}(1))\right\} \nonumber \\
&=\mathbb{E}_{D_e,g}\left\{f\left(g(\rho_e (x),X_e(1))\right)\right\} \nonumber \\
\textrm{($\delta$-A.C)} &\leq \delta + \mathbb{E}_{D_e} \left\{ \mathbb{E}_{g}\left\{f\left(g\left(\mathbb{E}_{D_e}\left\{\rho_e (x)\right\},X_e(1)\right)\right)\right\}\right\} \nonumber \\
\textrm{($\gamma$-B.E)} &< \delta +  \mathbb{E}_{D_e} \left\{ \mathbb{E}_{g}\left\{f\left(X_e(1)\right)\right\}\right\} - \gamma|\mathbb{E}_{D_e}\left\{\rho_e (x)\right\} - \rho_{eq}| \nonumber \\
\textrm{(\ref{eq:rho_e_lower})} &\leq \delta +  \mathbb{E}_{D_e} \left\{ \mathbb{E}_{g}\left\{f\left(X_e(1)\right)\right\}\right\} - \delta \nonumber \\
&< \rho_e^o(x) \label{eq:thm_error_lineB}
\end{align}
and correspondingly $\mathbb{E}_{D_e}\left\{\rho_{e+1}^o(x)\right\} > \rho_e^o(x)$ if $\rho_e^o(x)<\rho_{eq} - \frac{\delta}{\gamma \lambda}$. Thus $\rho_e^o(x)$ moves toward $\rho_{eq}$ in expectation if it is outside the interval $\left[\rho_{eq}- \frac{\delta}{\gamma \lambda},\rho_{eq}+ \frac{\delta}{\gamma \lambda}\right]$. 

Indeed, for $\rho_e^o(x) >\rho_{eq} + \frac{\delta}{\gamma \lambda} + \epsilon$ then this movement is bounded below; we have
\begin{equation}
\mathbb{E}_{D_e}\left\{\rho_e(x)\right\}  - \rho_{eq} \geq \lambda(\rho_e^o(x)-\rho_{eq}) > \frac{\delta}{\gamma} + \lambda \epsilon
\end{equation}
and, replacing the `$-\delta$' with `$-(\delta+\gamma\lambda\epsilon)$' in the second-to-last line of the argument in~\ref{eq:thm_error_lineB} we now have 
\begin{align}
\rho_e^o(x) - \mathbb{E}_{D_e}\{\rho_{e+1}^o(x)\} > \gamma \lambda \epsilon
\end{align}
Similarly, if $\rho_e^o(x) < \rho_{eq} - \frac{\delta}{\gamma \lambda} - \epsilon$, then $\mathbb{E}_{D_e}\{\rho_{e+1}^o(x)\} - \rho_e^o(x) > \gamma \lambda \epsilon$.

We now bound the maximum amount by which $\mathbb{E}_{D_e}\{\rho_{e+1}^o(x)\}$ can move towards $\rho_{eq}$. As in previous findings, it may `overshoot'. Firstly, suppose $\rho_e^o(x) (=\mathbb{E}\{f(X_e)\}) \geq I_1=\rho_{eq}-\frac{\delta}{\gamma \lambda}$. Now
\begin{align}
\inf_{e:\rho_e^o(x)\geq I_1} \mathbb{E}_{D_e}\{\rho_{e+1}^o(x)\} &= \inf_{e:\rho_e^o(x)\geq I_1} \mathbb{E}_{D_e,g}\left\{f\left(g(\rho_e (x),X_e(1))\right)\right\} \nonumber \\
\textrm{($\delta$-A.C)} &\geq -\delta + \inf_{e:\rho_e^o(x)\geq I_1} \mathbb{E}_{g}\left\{f\left(g(\mathbb{E}_{D_e}\{\rho_e (x)\},X_e(1))\right)\right\} \nonumber \\
\textrm{(\ref{eq:rho_e_lower})} &\geq -\delta + \inf_{\begin{smallmatrix} \rho \geq \rho_{eq} + \lambda(I_1-\rho_{eq}) \\ X:\mathbb{E}\{f(X)\} \geq I_1 \end{smallmatrix}} \mathbb{E}_{g}\left\{f\left(g(\rho,X)\right)\right\} \nonumber \\
&= J_1
\end{align}
and likewise, if $\rho_e^o(x) \leq I_2=\rho_{eq}+\frac{\delta}{\gamma \lambda}$, then $\sup_{e:\rho_e^o(x)\leq I_2} \mathbb{E}_{D_e}\{\rho_{e+1}^o(x)\} \leq J_2$. This completes the proof.

\end{proof}

\clearpage

\subsection{Theorem~\ref{thm:drift} (drift in $f$)}

\begin{reptheorem}{thm:drift}
Suppose that $\rho_e$, $G_e$ evolve as per algorithm~\ref{alg:main}, that the functions $f_e$ are $(q,\alpha)$-bounded about some function $f$, that the intervention $g$ is well-intentioned as per assumption~\ref{asm:rho_eq} at level $(q,\gamma)$ with respect to $f$, and assumptions~\ref{asm:g_closure} and \ref{asm:oracle} hold.  Defining 
\begin{align}
I_{\rho} &=\left[\rho_{eq}-2\frac{\alpha}{\gamma},\rho_{eq}+2\frac{\alpha}{\gamma}\right] \nonumber \\
S_{\inf} &= \left\{X \in \mathscr{X}:q\{\mathbb{E}\{f(X)\}\} \leq \alpha + q\left\{\rho_{eq} + \alpha\frac{1+\gamma}{\gamma}\right\}\right\} \nonumber \\
S_{\sup} &= \left\{X \in \mathscr{X}:q\{\mathbb{E}\{f(X)\}\} \geq -\alpha + q\left\{\rho_{eq} - \alpha\frac{1+\gamma}{\gamma}\right\}\right\} \nonumber \\
I_{\lim} &= \left[\inf_{X \in S_{\inf}} q^{-1}\left[-\alpha +  q\left\{\mathbb{E}\left\{f\left(g\left(\mathbb{E}\{f(X)\},X\right)\right)\right\}\right\}\right],  \sup_{X \in S_{\sup}} q^{-1}\left[\alpha +  q\left\{\mathbb{E}\left\{f\left(g\left(\mathbb{E}\{f(X)\},X\right)\right)\right\}\right\}\right]\right] \nonumber
\end{align}
we have that one of the following holds:
\begin{align}
\lim_{e \to \infty} \rho_e(x) &\in \left\{\min(I_{\rho}),\max(I_{\rho})\right\} \nonumber \\
\left[\lim \inf_{e \to \infty} \rho_e(x), \lim \sup_{e \to \infty} \rho_e(x)\right] &\subseteq I_{\lim} \nonumber
\end{align}
\end{reptheorem}

\begin{proof}

As usual, set $X_e(0)=x$ for all $e$. Suppose that for some $e>0$ we have
\begin{equation}
\mathbb{E}\{f_e(X_e)\}>\rho_{eq} + 2\frac{\alpha}{\gamma} + \epsilon = \max(I_{\rho}) + \epsilon \label{eq:Efe_epsilon}
\end{equation}
Then
\begin{align}
q\left\{\rho_{e+1}(x)\right\} &= q\left\{\mathbb{E}\left\{f_{e+1}\left(X_{e+1}(1)\right)\right\}\right\} \nonumber \\
&= q\left\{\mathbb{E}\left\{f_{e+1}\left(g\left(\rho_e(x),X_e(1)\right)\right)\right\}\right\} \nonumber \\
&= q\left\{\mathbb{E}\left\{f_{e+1}\left(g\left(\mathbb{E}\left\{f_e(X_e(1))\right\},X_e(1)\right)\right)\right\}\right\} \nonumber \\
\textrm{($(q,\alpha$)-B)} &\leq \alpha + q\left\{\mathbb{E}\left\{f\left(g\left(\mathbb{E}\left\{f_e(X_e(1))\right\},X_e(1)\right)\right)\right\}\right\} \nonumber \\
\textrm{(asm.~\ref{asm:rho_eq})} &\leq  \alpha + q\left\{\mathbb{E}\left\{f\left(X_e(1)\right)\right\}\right\} - \gamma\left|\mathbb{E}\left\{f_e(X_e(1))\right\} - \rho_{eq}\right| \nonumber \\
\textrm{($(q,\alpha$)-B)} &\leq 2\alpha + q\left\{\mathbb{E}\left\{f_e\left(X_e(1)\right)\right\}\right\} - \gamma\left|\mathbb{E}\left\{f_e(X_e(1))\right\} - \rho_{eq}\right| \nonumber \\
\textrm{\eqref{eq:Efe_epsilon}} &< q\left\{\mathbb{E}\left\{f_e\left(X_e(1)\right)\right\}\right\} - \gamma \epsilon \nonumber \\
&= q(\rho_e(x)) - \gamma\epsilon \nonumber
\end{align}
so $g$ moves $\rho_e(x)$ towards $\rho_{eq}$ if $\rho_e(x) \geq \max(I_{\rho}) + \epsilon$, and moves $q(\rho_e(x))$ towards $q(\rho_{eq})$ by at least $\gamma \epsilon$. A similar argument holds if $\rho_e(x) \leq \min(I_{\rho}) - \epsilon$. Thus if $\rho_e(x)>\max(I_{\rho})$ then the sequence $\{\rho_E(x),E>e\}$ either decreases until some $\rho_E$ is less than $\max(I_{\rho})$ or has limit $\max(I_{\rho})$, with the corresponding result if $\rho_e(x)<\max(I_{\rho})$.


The value $\lim\inf_{e \to \infty} \rho_e(x)$ is therefore bounded by the maximum `overshoot' downwards from when $\rho_e(x)$ starts within $I \cup [\rho_{eq},1]$. We firstly note that
\begin{align}
\mathbb{E}\{f_e(X)\} \in I_{\rho} &\Leftrightarrow \rho_{eq}-2\frac{\alpha}{\gamma} \leq \mathbb{E}\{f_e(X)\} \leq \rho_{eq}+2\frac{\alpha}{\gamma} \nonumber \\
&\Rightarrow -\alpha + \rho_{eq}-2\frac{\alpha}{\gamma} \leq \mathbb{E}\{f(X)\} \leq \alpha + \rho_{eq}+2\frac{\alpha}{\gamma} \mathbb{E}\{f(X)\} \nonumber \\
&\Rightarrow \mathbb{E}\{f(X)\} \in I_X
\end{align}
and hence
\begin{align}
\lim \sup_{e \to \infty} \rho_e(x) &\leq \sup_{e:\mathbb{E}\{f_e(x)\} \in (I_{\rho} \cup [0,\rho_{eq}])} \mathbb{E}\left\{f_{e+1}\left(g\left(\mathbb{E}\left\{f_e(x)\right\},X_e(1)\right)\right)\right\} \nonumber \\ 
&\leq \sup_{X \in \mathscr{X}:\mathbb{E}\{f_e(X)\} \in (I_{\rho} \cup [0,\rho_{eq}])} \mathbb{E}\left\{f_{e+1}\left(g\left(\mathbb{E}\{f(X)\},X\right)\right)\right\} \nonumber \\
&\leq \sup_{X \in \mathscr{X}:q\{\mathbb{E}\{f(X)\}\} \geq -\alpha + q\left\{\rho_{eq} - \alpha\frac{1+\gamma}{\gamma}\right\}} \mathbb{E}\left\{f_{e+1}\left(g\left(\mathbb{E}\{f(X)\},X\right)\right)\right\} \nonumber \\
&\leq \sup_{X \in \mathscr{X}:q\{\mathbb{E}\{f(X)\}\} \geq -\alpha + q\left\{\rho_{eq} - \alpha\frac{1+\gamma}{\gamma}\right\}} q^{-1}\left[\alpha +  q\left\{\mathbb{E}\left\{f\left(g\left(\mathbb{E}\{f(X)\},X\right)\right)\right\}\right\}\right] \nonumber
\end{align}
with a similar result for the limit inferior. 

\end{proof}

\clearpage

\subsection{Corollary~\ref{cor:shocks} (sudden drift in $f$)}

\begin{repcorollary}{cor:shocks}

Suppose that $\rho_e$, $G_e$ evolve as per algorithm~\ref{alg:main} and assumptions~\ref{asm:g_closure} and \ref{asm:oracle} hold. Suppose we have some sequence $E_1,E_2,\dots$ where $E_{i+1}-E_i > n_{\min}$ such that for $e \in [E_i,E_{i+1})$ all $f_e$ are $(q,\alpha)$-bounded around a function $f^i$ and that at each epoch $g$ is well-intentioned at level $(q,\gamma)$ with respect to $f^i$. Define $I_{\rho}^i$, $I_{\lim}^i$ as per theorem~\ref{thm:drift} with $f^i$ in place of $f$. Taking some $\epsilon_0>0$, if $n_{\lim}$ is sufficiently large (depending on $\epsilon_0$), there exists some $n_{\epsilon} < n_{\min}$ such that one of
\begin{align}
|\rho_e(x) - \min(I_{\rho}^i)| &< \epsilon_0 \nonumber \\ 
|\rho_e(x) - \max(I_{\rho}^i)| &< \epsilon_0 \nonumber \\ 
\rho_e(x) \in I_{\lim}^i \nonumber
\end{align}
holds for all $e$ with $E_i + n_i \leq e < E_{i+1}$

\end{repcorollary}

\begin{proof}
We appeal to a stronger form of theorem~\ref{thm:drift} evident from the proof: that rather than only the limit bounds of $\rho_{e}(x)$ being inside $I_{\lim}$, the sequence itself will eventually be constrained within $I_{\lim}$ or be within $\epsilon_0$ of one of $\min(I_{\rho})$, $\max(I_{\rho})$, after some number of epochs depending only on $\alpha$, $\lambda$ and $\epsilon_0$. As long as $n_{\min}$ exceeds this number of epochs, then the result follows.
\end{proof}

\clearpage

\section{Demonstration of results}
\label{supp_sec:theorem_simulations}

\sloppy
We give brief demonstrations of the main effects noted in theorems~\ref{thm:deterministic}, \ref{thm:probabilistic}, \ref{thm:convergence}, \ref{thm:errors} and \ref{thm:drift}, using worked examples. Annotated R scripts detailing these demonstrations are available at~\texttt{https://github.com/jamesliley/Stacked\_interventions}.

\begin{figure}[h]
\centering
\includegraphics[width=0.4\textwidth]{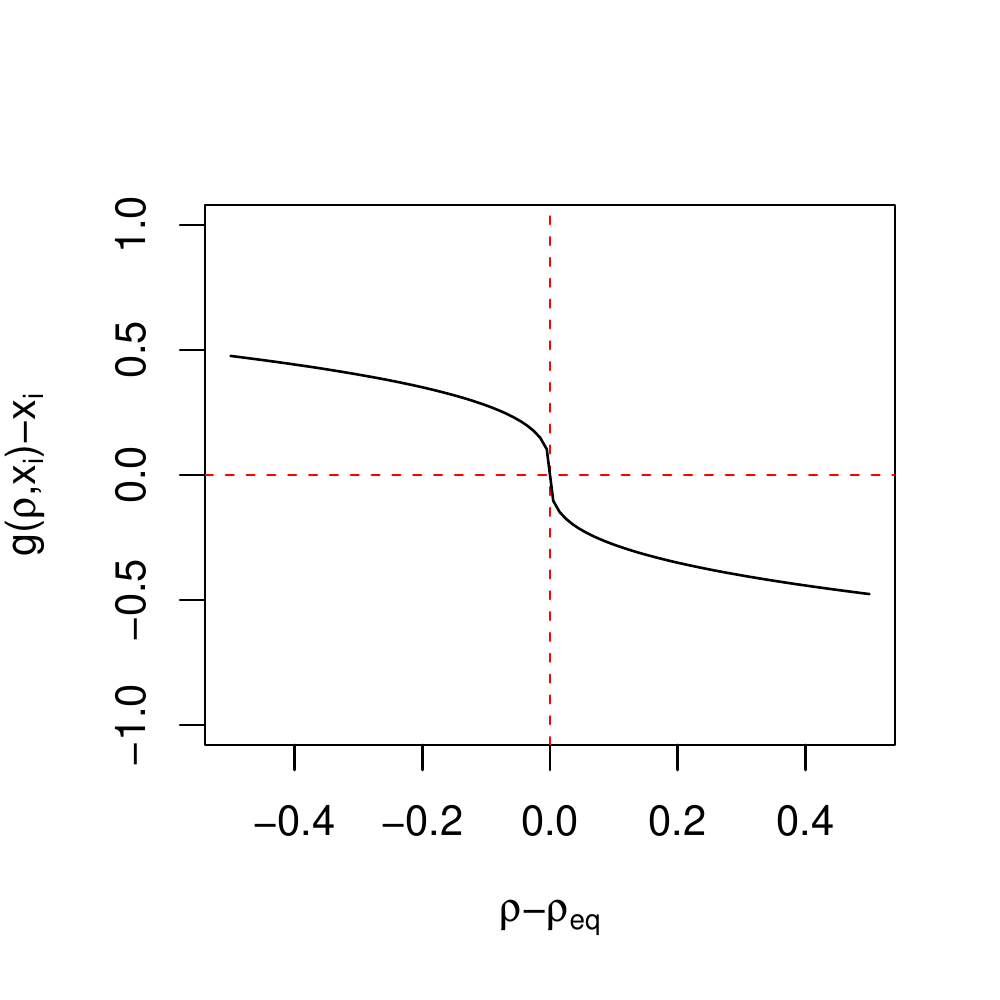}
\caption{Effect of $g(\rho,x)$ on $x$ for deterministic $g$. The effect of $g$ does not decrease linearly as $\rho \to \rho_{eq}$.}
\label{fig:g_example_deterministic}
\end{figure}

We considered $X$ comprising three real-valued covariates. We firstly set
\begin{equation}
f(x) = P(Y|X_e(1)=x) = \left(1+e^{-x^t \beta}\right)^{-1} \label{eq:f_definition_example}
\end{equation}
that is, a standard logistic link function. We used parameters $\beta=(\beta_1,\beta_2,\beta_3)$ chosen from $U(0,1)$ and set $\rho_{eq}=0.2$. We firstly considered an intervention function $g$ with the following deterministic effect, interpretable as `The effect of the agent's intervention on each covariate depends on the distance of $\rho$ from $\rho_{eq}$ in a non-linear way':
\begin{align}
g(\rho,x) &= x - \left(\rho-\rho_{eq}\right)^{\frac{1}{3}} 
\end{align}
We denote $q(x)=-\log\left(\frac{1}{x}-1\right)$. We have for all $\rho>\rho_{eq}$ and for all $x$, for some $\gamma>0$
\begin{equation}
q[f(x)]- q\left[f(g(\rho,x))\right]= x \beta - \left(x-(\rho-\rho_{eq})^{\frac{1}{3}}\begin{pmatrix} 1 \\ 1 \\ 1 \end{pmatrix}\right)^t \beta  \geq \gamma |\rho-\rho_{eq}|
\end{equation}
The effect of $g$ on $x$ is shown in figure~\ref{fig:g_example_deterministic}). We note that $\inf_{x \in \Xi:f(x)>\rho_{eq}} f\left(g(f(x),x)\right) < \rho_{eq}$ so $g$ can lead to `overshooting' $\rho_{eq}$: if $X_e(1)=x$ with $f(x)>\rho_{eq}$ for some $e$ (risk for a sample after intervention at epoch $e$ is larger than acceptable risk), then it is possible that $f(X_{e+1}(1))<\rho_{eq}$ (risk for a sample after intervention on epoch $e+1$ is lower than $\rho_{eq}$). Risk scores $\rho_e$ are become analytically complex for $e>1$ and hence we estimated these empirically. 
Assumptions~\ref{asm:g_closure}, \ref{asm:rho_eq}, and \ref{asm:no_drift} of theorem~\ref{thm:deterministic} readily hold, but assumption~\ref{asm:oracle} only holds approximately due to empirical approximation of $\rho_e$.

\begin{figure}
\begin{subfigure}{\textwidth}
  \begin{subfigure}{0.4\textwidth}
   \includegraphics[width=\textwidth]{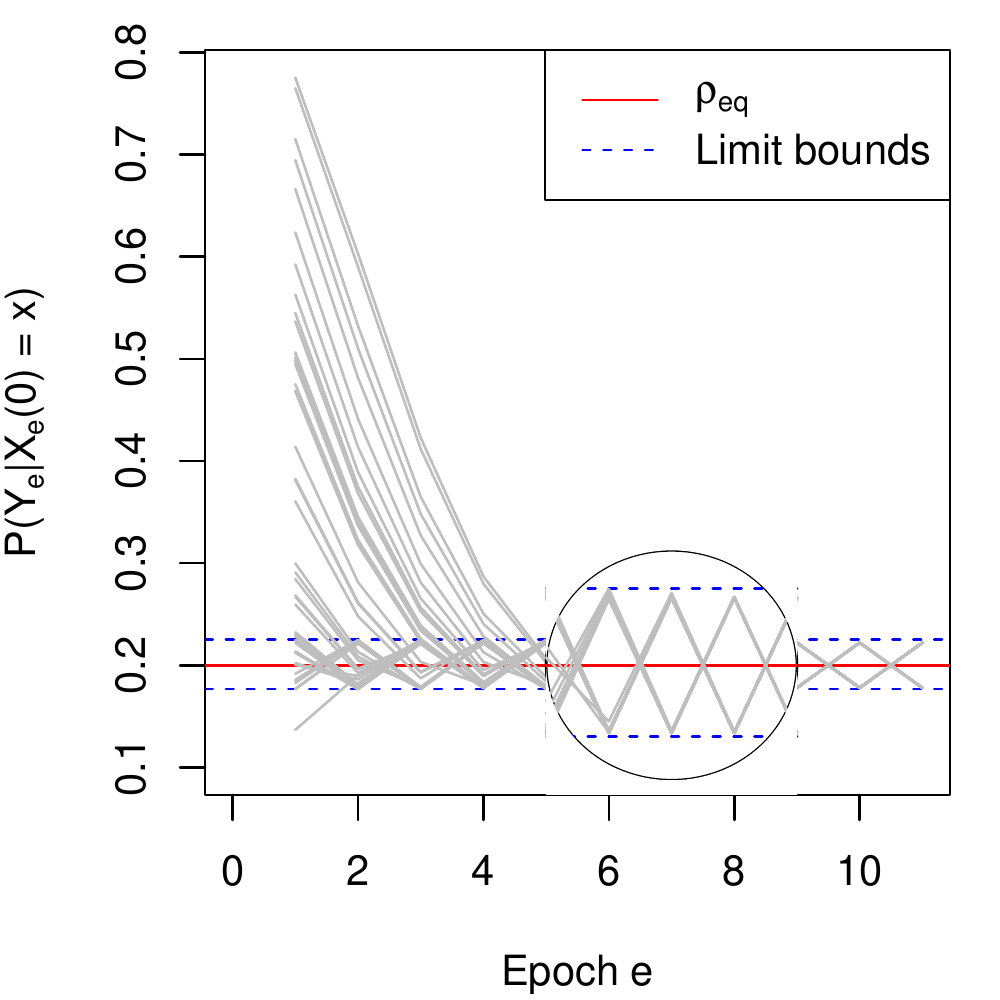}
   \caption{Theorem~\ref{thm:deterministic} illustration}
   \label{fig:thm_deterministic}
  \end{subfigure}
  \begin{subfigure}{0.4\textwidth}
   \includegraphics[width=\textwidth]{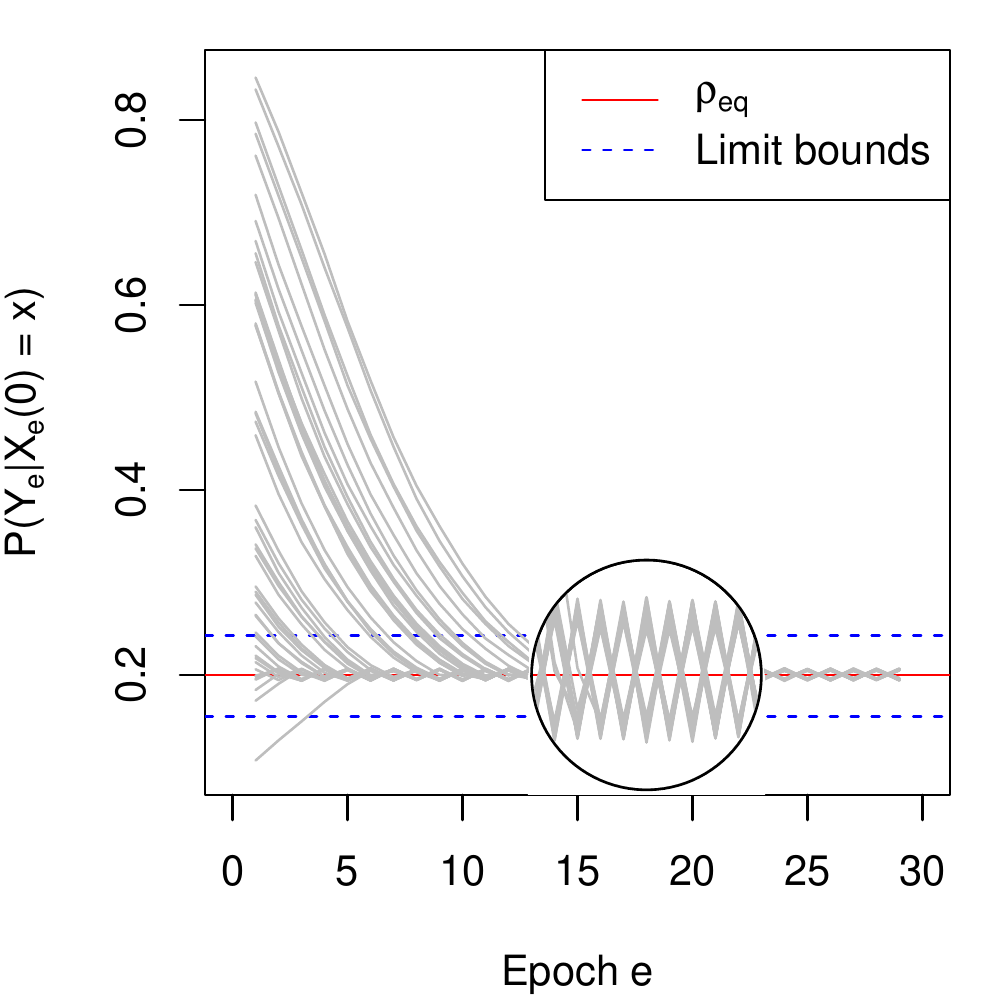}
   \caption{Theorem~\ref{thm:probabilistic} illustration}
   \label{fig:thm_probabilistic}
  \end{subfigure}
\end{subfigure}
\begin{subfigure}{\textwidth}
  \begin{subfigure}{0.4\textwidth}
   \includegraphics[width=\textwidth]{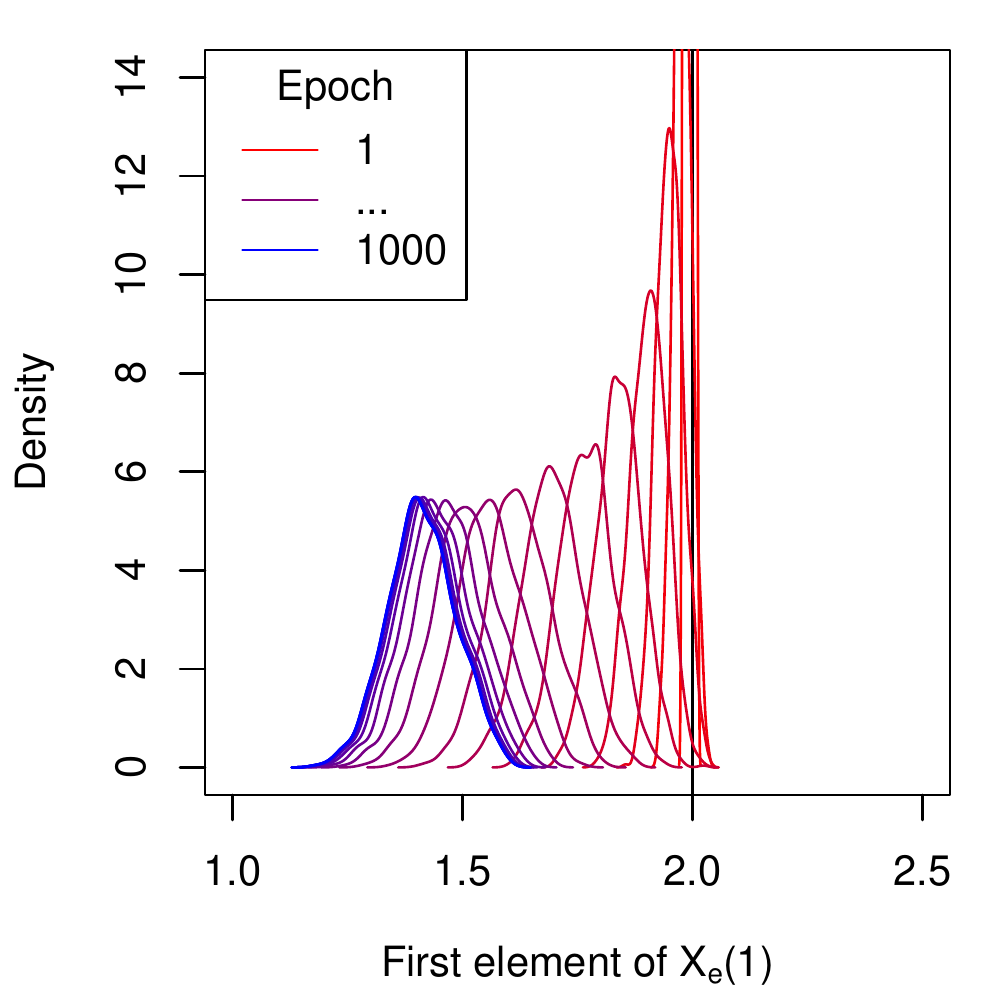}
   \caption{Theorem~\ref{thm:convergence}: convergence}
   \label{fig:thm_convergence1}
  \end{subfigure}
  \begin{subfigure}{0.4\textwidth}
   \includegraphics[width=\textwidth]{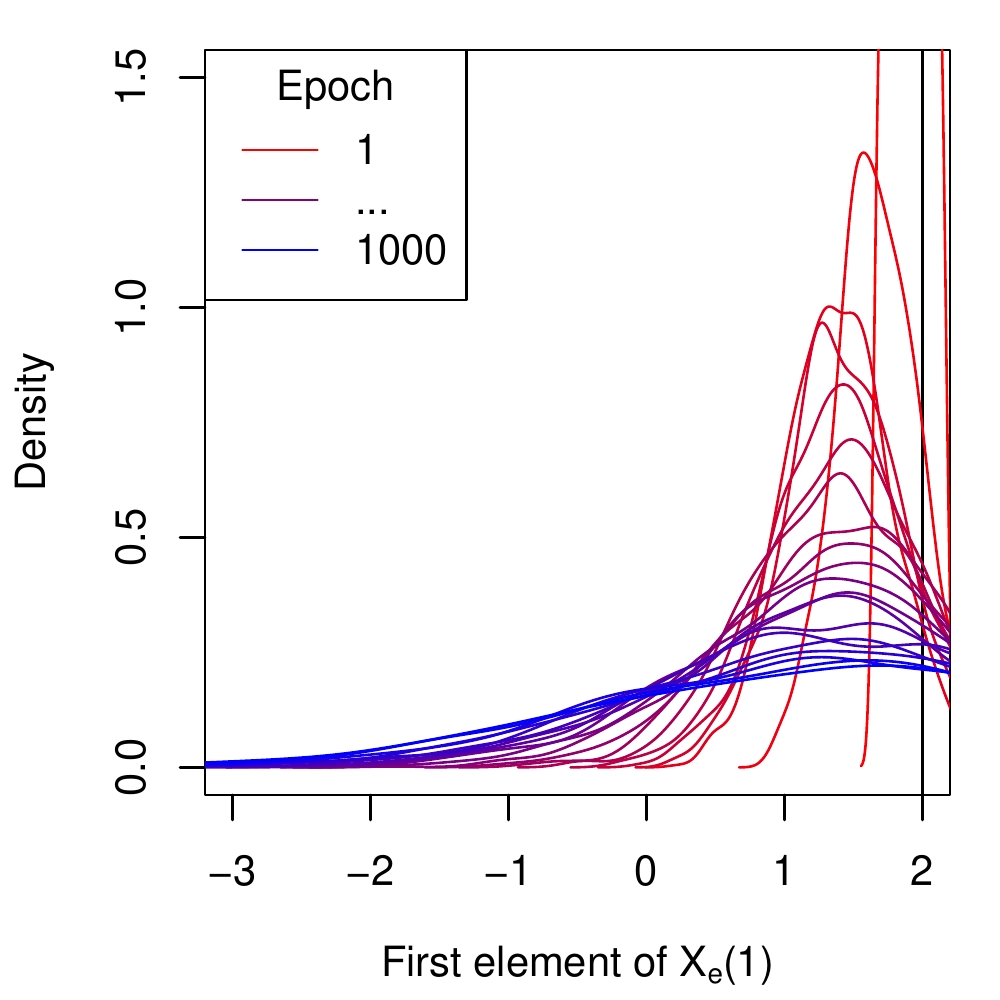}
   \caption{Theorem~\ref{thm:convergence}: non-convergence}
   \label{fig:thm_convergence2}
  \end{subfigure}
\end{subfigure}
\caption{Illustrations of theorems~\ref{thm:deterministic}, \ref{thm:probabilistic}, and \ref{thm:convergence}. In panels~\ref{fig:thm_deterministic} and  \ref{fig:thm_probabilistic}, the Y-axis plots $\rho_e(x)$, or the probability of the event $Y_e$ in epoch $e$ for a sample which started with covariates $x$. Fifty values of starting covariates are shown, plotted as grey lines. Panels~\ref{fig:thm_convergence1} and~\ref{fig:thm_convergence2} show density estimates for the first element of $X_e(1)$ with changing $e$, given a fixed value $X_e(0)=x$. Epochs run from $e=1$ to $e=1000$ but for clarity not all epochs are shown. In panel~\ref{fig:thm_convergence1}, the distributions of the first element of $X_e(1)$ can be seen to converge. In panel~\ref{fig:thm_convergence2}, they do not.
}
\label{fig:simulation_theorems1}
\end{figure}

Figure~\ref{fig:thm_deterministic} shows the eventual containment of $\rho_e$ in the limit bounds around $\rho_{eq}$, starting from a range of initial values of $X_e(0)$.

\begin{figure}[h]
\centering
\includegraphics[width=0.4\textwidth]{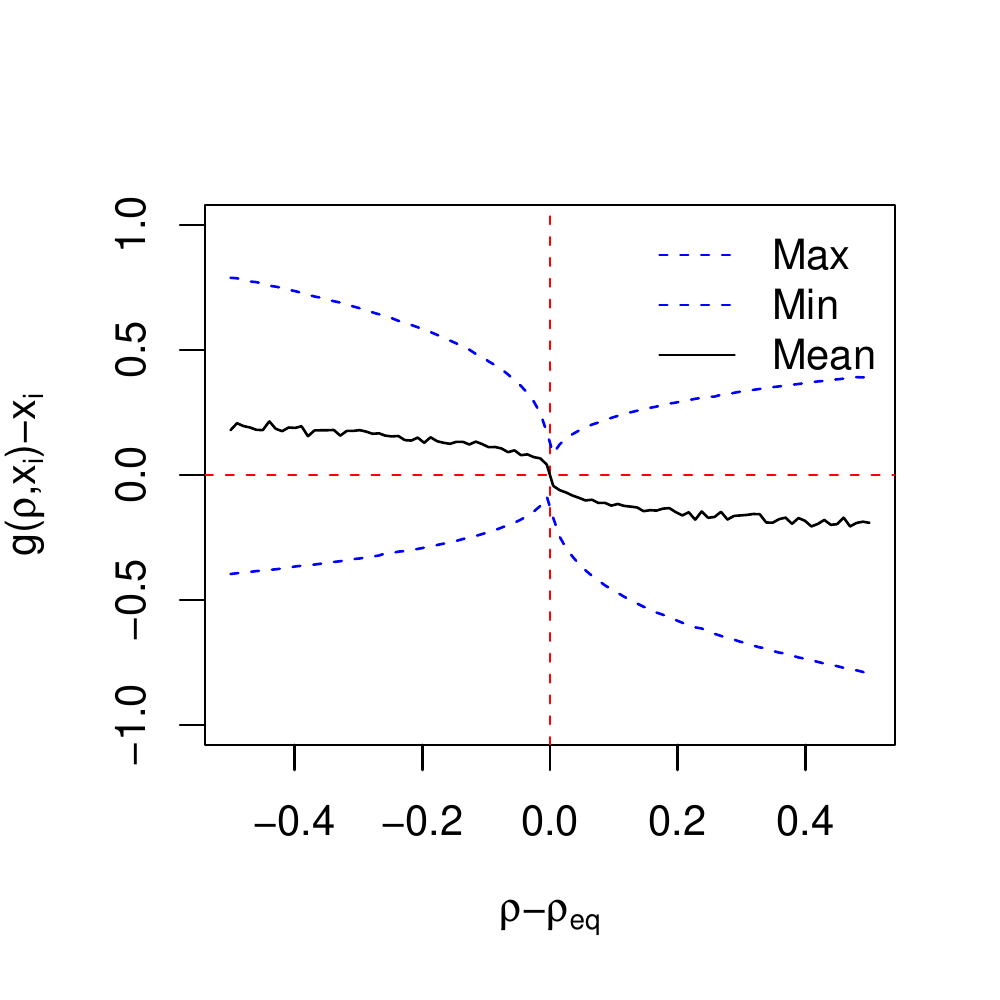}
\caption{Effect of $g(\rho,x)$ for probabilistic $g$. The effect of $g$ is random: it moves $x$ by some amount which is uniformly-distributed, and can move $x$ in the `wrong' direction. The expected (mean) effect of $g$ does not decrease linearly as $\rho \to \rho_{eq}$.}
\label{fig:g_example_probabilistic}
\end{figure}

For theorem~\ref{thm:probabilistic}, we used a similar formulation  of $g$ but with a random element:
\begin{align}
g(\rho,x) &= x - (\rho-\rho_{eq})^{\frac{1}{3}} U\left(-\frac{1}{2},1\right) \label{eq:g_thm_probabilistic} 
\end{align}
where $U(a,b)$ is a uniformly-distributed random variable with range $[a,b]$. Figure~\ref{fig:g_example_probabilistic} shows the maximum, minimum and expected effect of $g(\rho,\cdot)$ on $x$ according to $\rho-\rho_{eq}$. We define $\mathscr{X}$ as the set of finite sums of uniformly-distributed random variables (which is closed under $g$) and similarly to the previous example, with $q(x)=-\log\left(\frac{1}{x}-1\right)$, for $X \in \mathscr{X}$ we have
\begin{equation}
q\left[\mathbb{E}\{f(X)\}\right]- q\left[\mathbb{E}\left\{f(g(\rho,X))\right\}\right]= \mathbb{E}\{X\}^t \beta - \left(\mathbb{E}\{X\}-\frac{1}{4}(\rho-\rho_{eq})^{\frac{1}{3}}\begin{pmatrix} 1 \\ 1 \\ 1 \end{pmatrix}\right)^t \beta  \geq \gamma |\rho-\rho_{eq}|
\end{equation}
Similarly, the intervention $g$ can lead to `overshooting' $\rho_{eq}$. Moreover, $g$ can `move' $x$ in the `wrong' direction, in that even for $\rho>\rho_{eq}$
\begin{equation}
P\left(f(g(\rho,x))>f(x)\right) =\frac{1}{3}> 0
\end{equation}
The probabilistic containment of $P(Y_e|X_e(0)=x)$ in limit bounds for a range of values of $x$ is shown in figure~\ref{fig:thm_probabilistic}. Note (from magnified portion) that there is more variation in the value of $P(Y_e|X_e(0)=x)$ for high $e$ in figure~\ref{thm:probabilistic} than in figure~\ref{thm:deterministic}. As for the previous example, assumption~\ref{asm:oracle} only approximately holds (for convenience's sake).

To demonstrate theorem~\ref{thm:convergence}, we considered two different versions of $g$, one of which led to convergence in distribution of $X_e$, and one which did not. Firstly, we used:
\begin{align}
g(\rho,x) &= x - \frac{1}{10}(\rho-\rho_{eq})U\left(-\frac{1}{2},1\right) \label{eq:g_thm3_def1}
\end{align}
With $q(\cdot)$ and $\mathscr{X}$ as for the previous example, we have
\begin{equation}
\left|\frac{q\left\{\mathbb{E}\left[f(X^1)\right]\right\}-q\{\rho_{eq}\}}{q\left\{\mathbb{E}\left[f(X)\right]\right\}-q\{\rho_{eq}\}}\right|=
\left|\frac{\mathbb{E}\{X^1\}^t \beta -q\{\rho_{eq}\}}{\mathbb{E}\{X\}^t \beta-q\{\rho_{eq}\}}\right| = \left|1 - \frac{1}{40}\frac{(\rho-\rho_{eq})|\beta|_1}{\mathbb{E}\{X\}^t \beta-q\{\rho_{eq}\}}\right| \label{eq:thm3_qbound}
\end{equation}
We used $\rho_{eq}=1/2$ in this case. Since $X_e(1)$ is a sum of uniform random variables it is symmetric and unimodal; hence $\textrm{sign}(f(E\{X\})-\rho_{eq}) = \textrm{sign}(\mathbb{E}\{f(X)\}-\rho_{eq})$. From the form of~\ref{eq:g_thm3_def1} it is clear that $\rho_e(x)=$ always gets closer to $\rho_{eq}$ as $e$ increases, and hence so does $\mathbb{E}\{f(X_e(1))\}$, so $\mathbb{E}(X_e(1))$ always decreases in magnitude. Thus the denominator of~\ref{eq:thm3_qbound} is bounded above and the quantity is bounded above by $1-\epsilon$ for some $\epsilon$ dependent on $x$. The quantity $\mathbb{E}\{X_e(1)\}$ can also be shown to be bounded for $\rho_{eq} \neq 1/2$, although details are more complex.

For $\rho = \mathbb{E}\{f(x)\}>\rho_{eq}$, we have
\begin{equation}
g(\rho,x)-x =  -\frac{1}{10}(\rho-\rho_{eq})U \defeq Z
\end{equation}
and 
\begin{align}
\mathbb{E}\{|Z|\} &=  -\frac{1}{40}(\rho-\rho_{eq}) = O(|\rho-\rho_{eq}|) \nonumber \\
\textrm{var}\{|Z|\} &=  -\frac{3}{24}(\rho-\rho_{eq}) = O(|\rho-\rho_{eq}|) \nonumber
\end{align}
The convergence in distribution of the first element of $X_e$ can be seen in figure~\ref{fig:thm_convergence1}. Secondly, we used a slightly different version of $g$:
\begin{align}
g(\rho,x) &= x - \frac{1}{10}(\rho-\rho_{eq})^{\frac{1}{3}}U\left(-\frac{1}{2},1\right) \nonumber
\end{align}
for which
\begin{equation}
\mathbb{E}\{|Z|\} =  -\frac{1}{40}(\rho-\rho_{eq})^{\frac{1}{3}}  \neq O(|\rho-\rho_{eq}|) \nonumber
\end{equation}
The non-convergence in distribution of the first element of $X_e$ can be seen in figure~\ref{fig:thm_convergence2}.

To demonstrate theorem~\ref{thm:errors}, we simulated $D_e$ by sampled 100 training samples of independent and identically distributed $X_e(0)\sim N(0,1)$ for each epoch $e$. We then followed algorithm~\ref{alg:main}, training $\rho_e$ to $D_e$ at epoch $e$ using a logistic model, leading to a system of $f$, $g$ and $\rho_e$ which was $\delta$-almost convex and had a $\lambda$-order effect with $\delta \approx 0.015$ and $\lambda=1/4$, and $\gamma \approx 1$ in assumption~\ref{asm:rho_eq}. We used $g(\rho,x)$ as per~\ref{eq:g_thm_probabilistic}.

\begin{figure}
\begin{subfigure}{\textwidth}
  \begin{subfigure}{0.4\textwidth}
   \includegraphics[width=\textwidth]{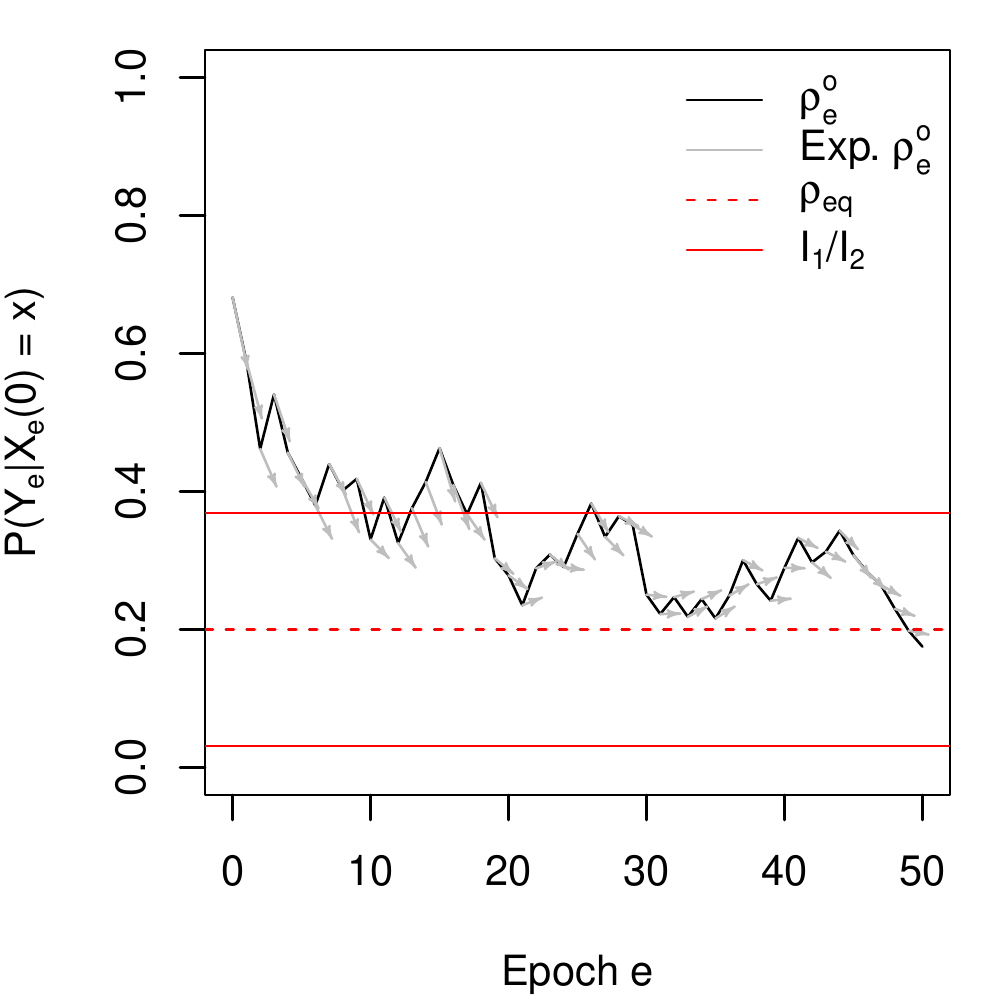}
   \caption{Theorem~\ref{thm:errors} illustration}
   \label{fig:thm_errors}
  \end{subfigure}
  \begin{subfigure}{0.4\textwidth}
   \includegraphics[width=\textwidth]{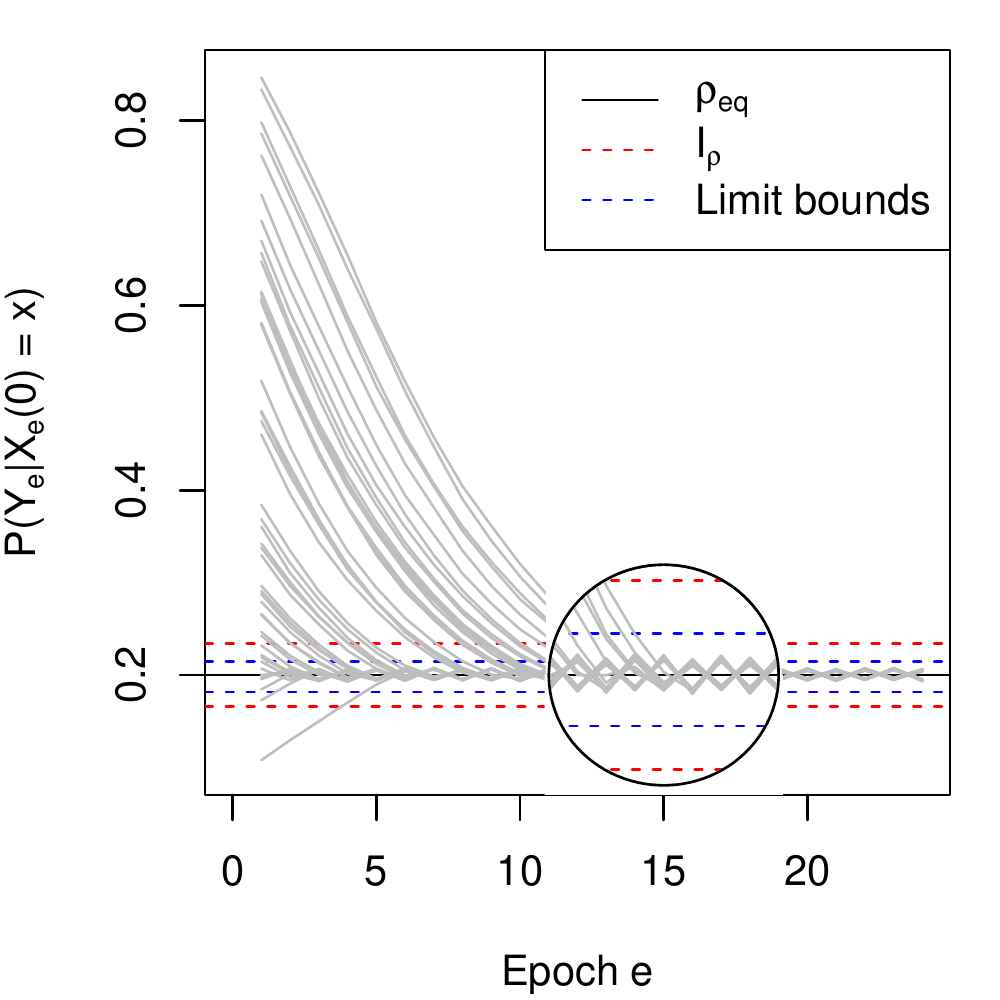}
   \caption{Theorem~\ref{thm:drift} illustration}
   \label{fig:thm_drift}
  \end{subfigure}
 \end{subfigure}
\begin{subfigure}{0.8\textwidth}
   \includegraphics[width=\textwidth]{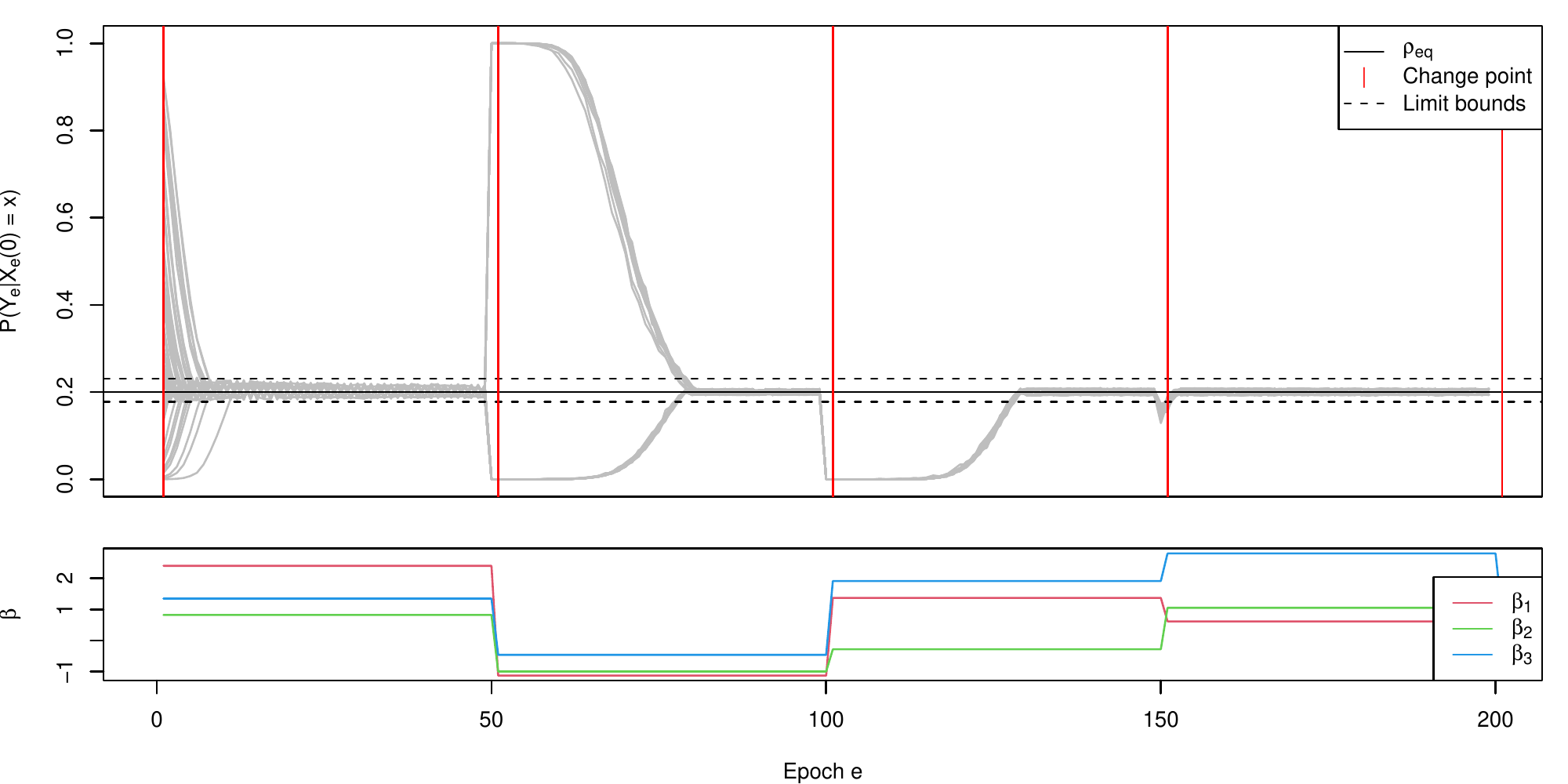}
   \caption{Corollary~\ref{cor:shocks} illustration}
   \label{fig:cor_shocks}
\end{subfigure}
\caption{Illustrations of theorems~\ref{thm:errors}, \ref{thm:drift} and corollary~\ref{cor:shocks}. In all panels except the bottom-most, the Y-axis plots $\rho_e^o(x)$,  or the probability of the event $Y_e$ in epoch $e$ for a sample which started with covariates $x$. In panel~\ref{fig:thm_errors} this is in contrast to the estimated probability $\rho_e(x)$. Gray arrows in panel~\ref{fig:thm_errors} indicate the expected movement of $\rho_e^o(x)$: that is, where we expect expected risk $\rho_{e+1}(x)$ will be at the next epoch. When $\rho_e(x)$ is outside the solid red lines, the expected movement is always back toward the red lines. In panels~\ref{fig:thm_drift} and~\ref{fig:cor_shocks}, the function $f$ (parametrised by $\beta$) changes randomly epoch-to-epoch: by small amounts each epoch around a constant function in figure~\ref{fig:thm_drift} and by a large amount periodically (solid red vertical lines) in panel~\ref{fig:cor_shocks}, for which changes in $\beta$ are shown in the bottom-most panel. Fifty values of starting covariates are shown, plotted as grey lines.
}
\label{fig:simulation_theorems}
\end{figure}

The behaviour of $\rho_e^o(x)$ can be seen to be `attracted' in expectation by the interval $[I_1,I_2]$ as shown in figure~\ref{fig:thm_errors}. When $\rho_e^o(x) = P(Y_x|X_e(0)=x)$ is outside $[I_1,I_2]$ it moves towards the interval in expectation.


For theorem~\ref{thm:drift}, we used $g$ as per equation~\ref{eq:g_thm_probabilistic}, but allowed $f_e$ to vary with $e$ such that for all $x\in \mathbb{R}^3$:
\begin{equation}
|q(f_e(x))-q(f(x))| \leq \frac{1}{200}
\end{equation}
Using $q(x)=-\log\left(\frac{1}{x}-1\right)$, we have that assumption~\ref{asm:rho_eq} holds at level $(q,\gamma)$ with $\gamma \approx 0.3$. The eventual containment of $P(Y_e|X_e(0)=x)$ is shown in figure~\ref{fig:thm_drift}. 

Finally, to demonstrate corollary~\ref{cor:shocks}, we designated a series of functions $f^ii$ defined as per equation~\ref{eq:f_definition_example} with $\beta^ii$ in the place of $\beta$. We considered 200 epochs, with three evenly-spaced change points $E_i$. At each change point, $\beta^i$ changed to three new values chosen independently from $U[-2,2]$. 

In contrast to previous examples, we constricted the range of each element of $X_e(1)$ to $[-5,5]$. Had we not done this, since the effect of `stacking' interventions $g$ is to move each element of $\mathbb{E}\{X\}$ in proportion to $(\rho-\rho_{eq})^{1/3}$, the aggregate effect of acting on a large number of risk scores fitted during a period with different $f$ would take many epochs to `correct'. We defined $g$ during the time period for which $f=f^i$ as:
\begin{equation}
g(\rho,x_j)=\textrm{clamp}\left(x_j - \textrm{sign}(\beta^i_j)(\rho-\rho_{eq})^{\frac{1}{3}},[-4,4]\right)
\end{equation}
where $j$ indexes elements of $\beta$ and $x$ and $\textrm{clamp}(\cdot,[a,b])=\min(b,\max(a,\cdot))$. This ensures that $g$ is well-intentioned with respect to $f^i$, with $q$ as above.

The recovery of $\rho_e(x)$ following change points can be seen in figure~\ref{fig:cor_shocks}.

\clearpage

\section{Details of simulation}
\label{supp_sec:example_details}

Our outcome is `use of secondary healthcare (e.g., hospital services) during the upcoming year'. Although in some sense we want to reduce risk of secondary healthcare use to as low a level as possible for each patient, we have only limited resources and must distribute them appropriately, and excessive action to reduce risk is unwarranted; there is little justification to make potentially onerous lifestyle adjustments for a patient at no higher risk than the population mean.

We presume that once a year health practitioners record the following:
\begin{enumerate}[leftmargin=4cm]
\item[Demographic details]: age, sex, socioeconomic deprivation quintile
\item[Medical history]: rating of severity of medical history
\item[Lifestyle]: rating of diet, smoking status, and alcohol usage
\item[Medications]: whether or not on one of ten medications which reduce adverse outcome risk
\end{enumerate}

We model deprivation on an integral five point scale (analogous to deprivation quintiles, eg~\citep{mclennan19}), and medical history on a ten-point scale, with a higher value indicating more extensive history. We model alcohol use and diet as real values in $[0,10]$, with higher values indicating higher-risk activity. We modelled sex, smoking status and status of each of ten medications as binary variables. We sampled covariates to roughly resemble values observable in a real population, specified in table~\ref{supp_tab:demographics}, for 10,000 random `individuals'.

We modelled `ground-truth' risk as dependent on covariate values through a generalised linear model with logistic link. We chose coefficients such that increased age, more extensive medical history, positive smoking status, worse diet, worse alcohol use, and `sex=1' conferred increased risk. We chose coefficients for the ten medications such that medications conferred either increased or decreased risk. Coefficients are specified in supplementary  table~\ref{supp_tab:demographics}. We considered nine categories of individuals defined by age and medical history (but not sex or deprivation level). In each category, we defined a separate intervention scheme, resulting in a separate `acceptable' level of risk. 

We presumed that `practitioners' could intervene as described above. We also presumed that practitioners would direct attention away from individuals with low risk scores, with the effect that covariates for such intervals would tend to worsen instead, with the effect increasing the lower the risk score. We designated both of these effects to be random (that is, corresponding to random-valued $g$ rather than deterministic). For diet and alcohol, the post-intervention value was the pre-intervention value changed by a uniformly-distributed random variable with constant variance and mean depending on the risk score. For smoking and medications, the post-intervention value was a Bernoulli random variable with mean depending on the previous value and score. An exact specification of the intervention is given in Supplementary table~\ref{supp_tab:intervention}, illustrated in Supplementary Figure~\ref{supp_fig:simulation_details}. 

After each intervention, a score was refitted to the pre-intervention covariates and post-intervention outcome using a random forest with 500 trees (noting that this differs from the ground-truth function for risk). The conditions for theorem~\ref{thm:probabilistic} were roughly satisfied: the intervention always `moved' the expectated risk towards the appropriate $\rho_{eq}$ value, but since modifiable covariates (alcohol, diet, smoking, medications) can reach minimum values, the magnitude of this movement is not necessarily bounded below (contadicting the `$\gamma$' condition in assumption~\ref{asm:rho_eq}). In addition, the fitted risk score $\rho$ was imperfect (not an oracle). However, these violations were sufficiently subtle that the principal effect illustrated in the theorem can be observed.

\begin{table}
\begin{tabular}{l|l|lll}
 & Variable & Distribution & Range (type) & Coefficient  \\
\hline
Fixed & Age & $P_{age}(x) \propto 1-\frac{4}{5}\frac{x}{100}$ & 0-99 (int) & $5 \times 10^{-3}$ \\
 & Sex & $P_{sex}(x) = \frac{1}{2}$ & 0-1 (bin) & 0.25 \\
 & Deprivation & $P_{dep.}(x) = \frac{1}{5}$ & 1-5 (int) & 0.1 \\
 & Medical history & $P_{med.hx}(x) = \frac{1}{10}$ & 1-10 (int) & 0.05 \\
 \hline
Modifiable & Smoking & $P_{smoking}(x)=\frac{3}{10}$ & 0-1 (bin) & 0.2 \\
& Diet & $f_{diet}(x) \propto 1$ & 0-10 (real) & 0.1 \\
& Alcohol & $f_{alcohol}(x) \propto 1$ & 0-10 (real) & 0.2 \\
& Drugs (1-10) & $P_{drug i}(x) = \alpha_i$, $0<\alpha_i<\frac{1}{2}$ & 0-1 (bin) & $\sim N(0,0.1^2)$
\end{tabular}
\caption{Distribution of variables used in simulation, and coefficients in ground truth model. Values for each individual were sampled independently. Although it would be more appropriate to allow dependence, this is not important for our simulation. The constant term in the ground-truth logistic regression model was -3.32. $N(\mu,\sigma^2$ denotes the Gaussian distribution with mean $\mu$ and variance $\sigma^2$}
\label{supp_tab:demographics}
\end{table}

\begin{table}
\begin{tabular}{l|ll}
Variable & New value  \\
\hline
Diet, alcohol: &  $U\left(x-\frac{7}{2}(\rho-\rho_{eq}),x+\frac{1}{2}(\rho-\rho_{eq}\right)$\\
Smoking, drugs (1:10): &  $\textrm{Bern}\left(\begin{cases} 2\left(1- \textrm{logistic}\left(\frac{1}{2} (\rho-\rho_{eq})\right)\right) &\textrm{ if } \rho>\rho_{eq}, x=1 \\ 
0 &\textrm{ if } \rho>\rho_{eq}, x=0 \\
2 \textrm{logistic}\left(\frac{1}{2}(\rho_{eq}-\rho)\right) - 1 &\textrm{ if } \rho \leq \rho_{eq}, x=0 \\
1 &\textrm{ if } \rho \leq \rho_{eq}, x=1 \end{cases}\right)$ \\
\end{tabular}
\caption{Intervention effects. The original value of the covariate is denoted $x$, the risk score $\rho$ and the `acceptable risk' $\rho_{eq}$. $U(a,b)$ denotes the uniform distribution on $[a,b]$; $\textrm{Bern}(p)$ the Bernoulli distribution with mean $p$; and $\textrm{logistic}(x)=(1 + e^{-x})^{-1}$. Note that interventions can move $x$ in such a way that $\rho$ moves away from $\rho_{eq}$. }
\label{supp_tab:intervention}
\end{table}

\begin{figure}
  \begin{subfigure}{0.5\textwidth}
   \includegraphics[width=\textwidth]{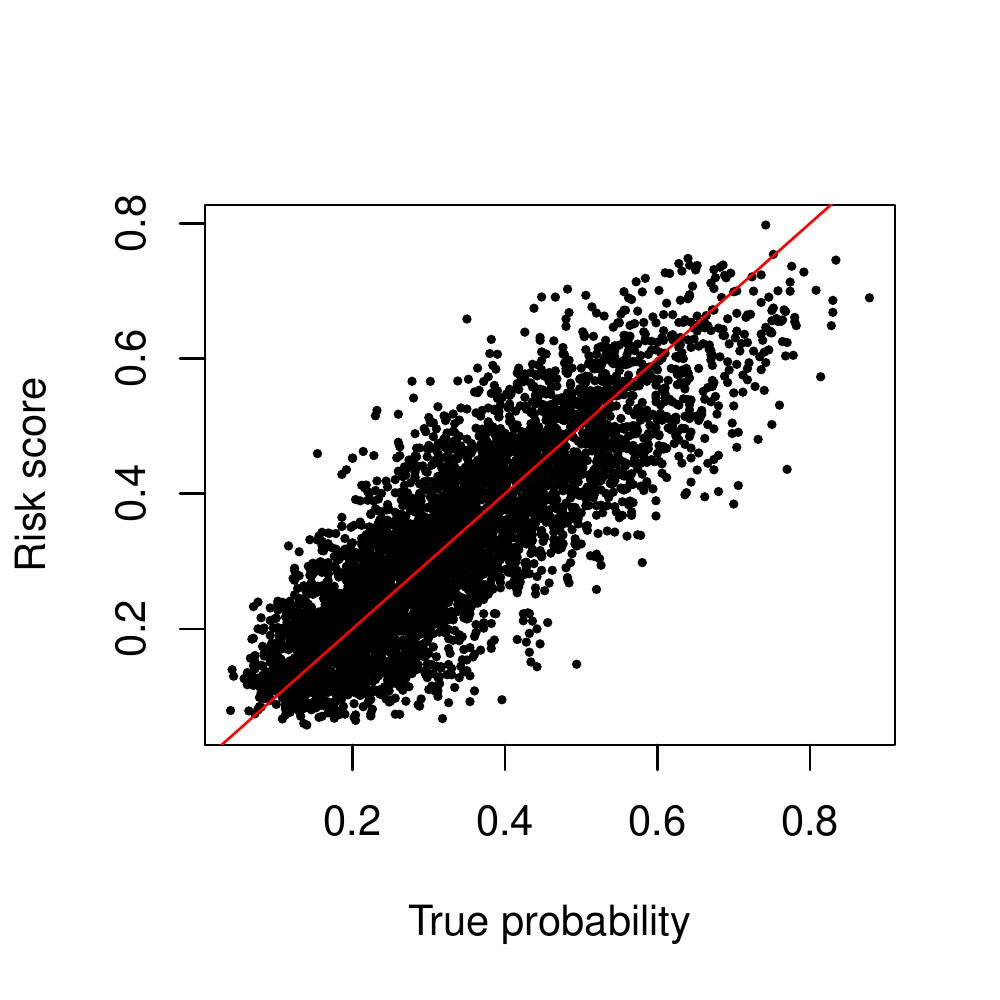}
   \caption{Fitted and true risks}
   \label{fig:fitted_vs_true}
  \end{subfigure}
  \begin{subfigure}{0.5\textwidth}
   \includegraphics[width=\textwidth]{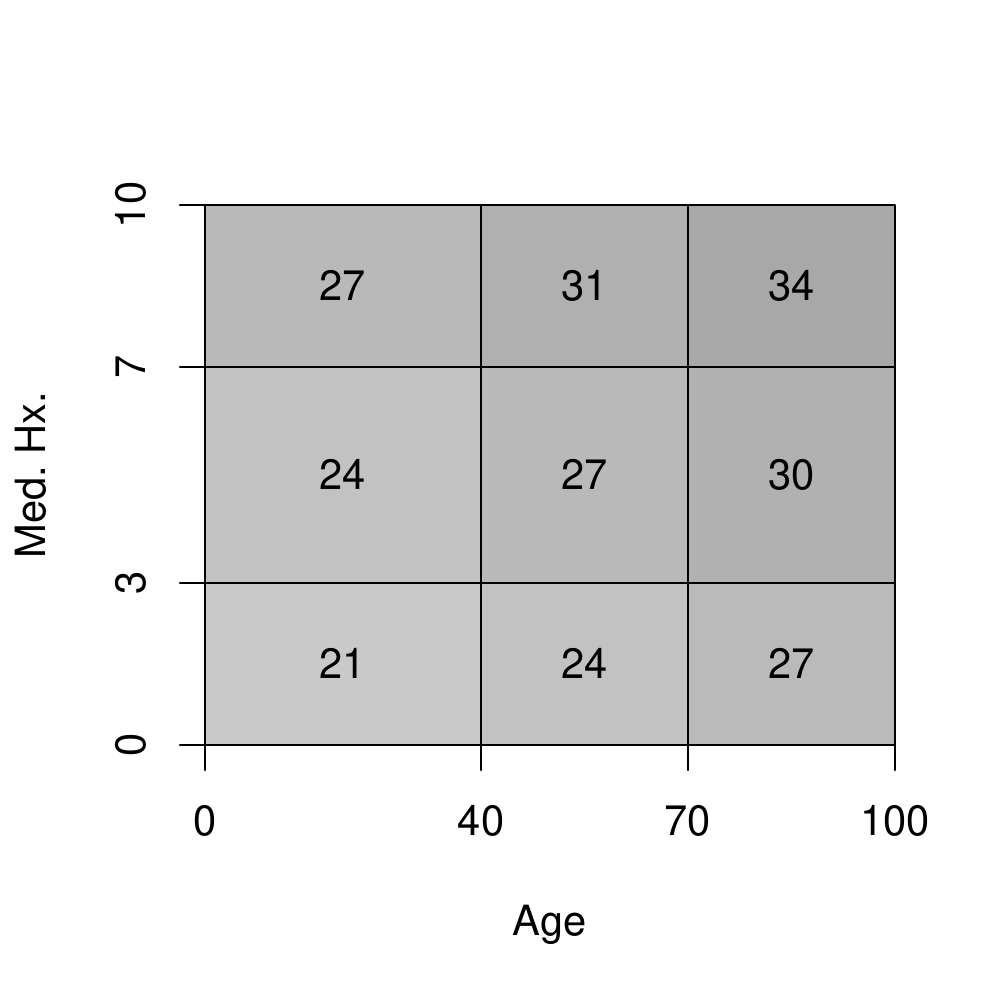}
   \caption{Values of $\rho_{eq}$ by cohort}
   \label{fig:rho_eq_by_cohort}
  \end{subfigure}
  \begin{subfigure}{0.5\textwidth}
   \includegraphics[width=\textwidth]{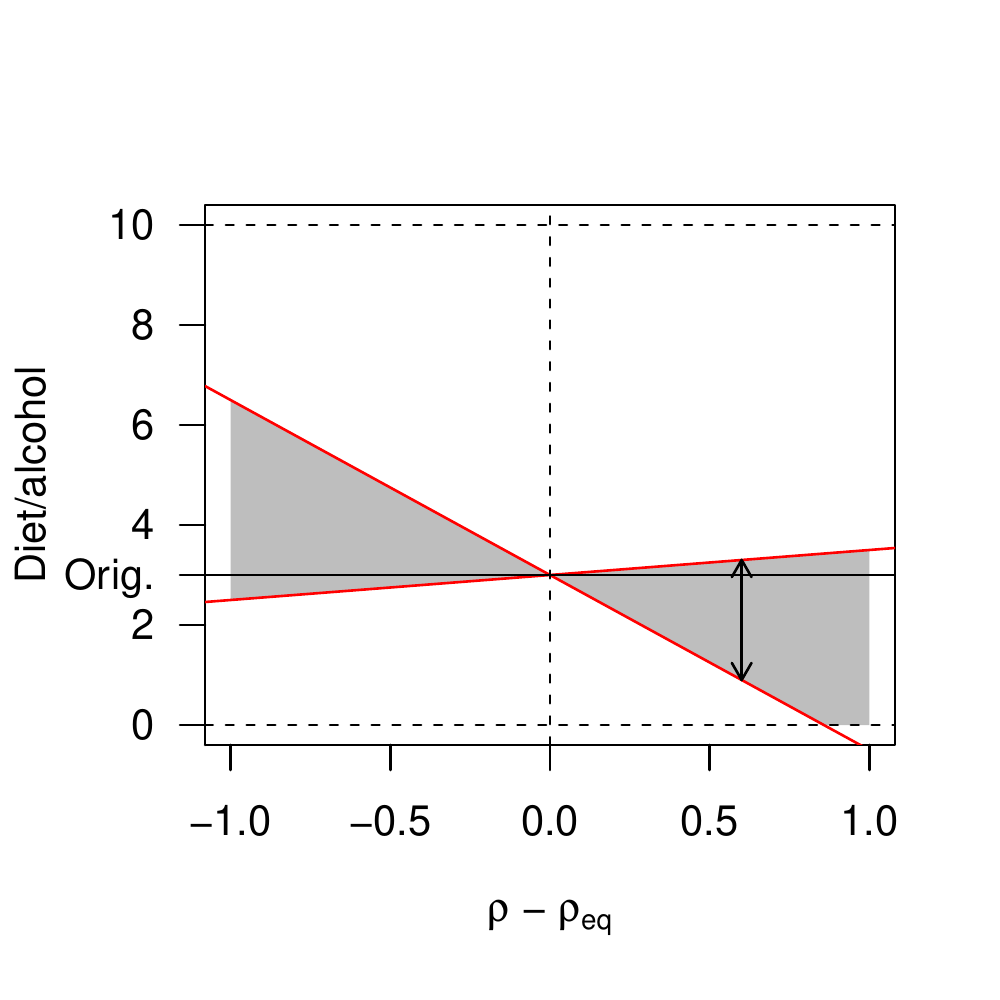}
   \caption{Intervention on continuous covariates}
   \label{fig:intervention_continuous}
  \end{subfigure}
  \begin{subfigure}{0.5\textwidth}
   \includegraphics[width=\textwidth]{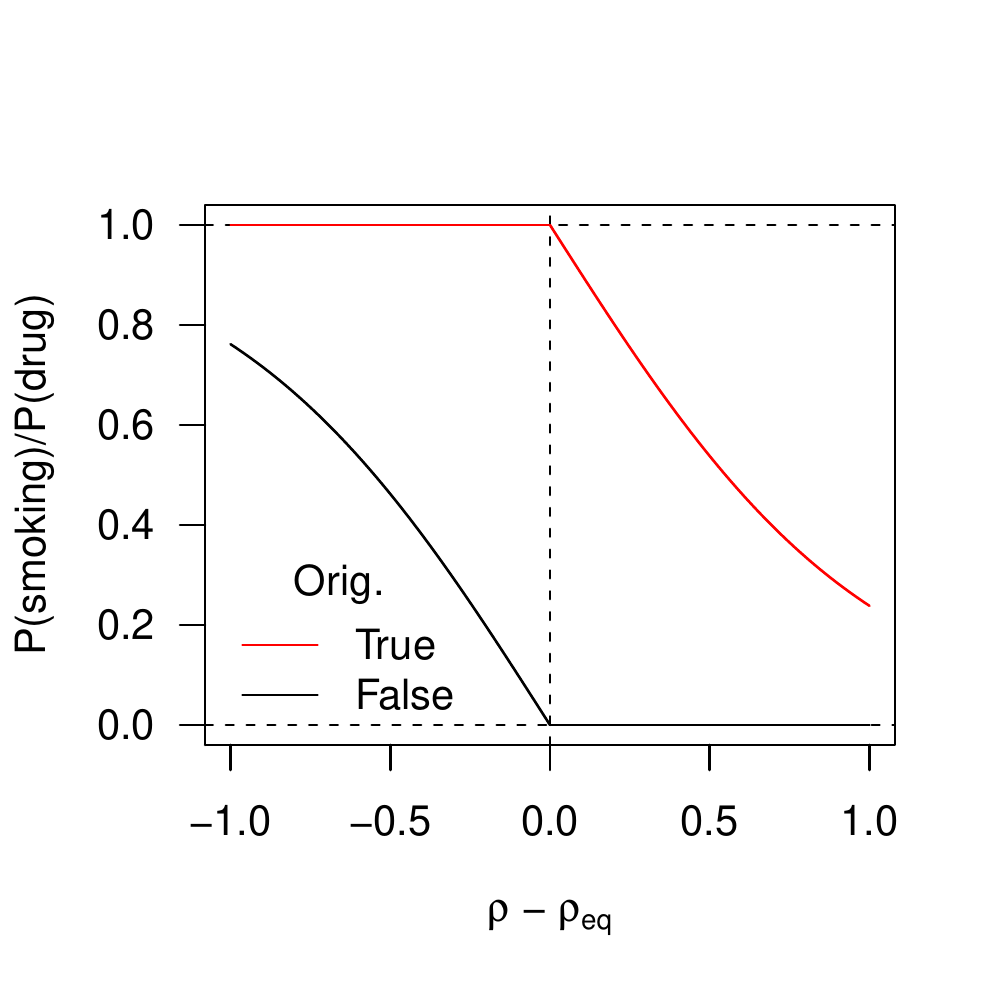}
   \caption{Intervention on binary covariates}
   \label{fig:intervention_binary}
  \end{subfigure}
\caption{Panel~\ref{fig:fitted_vs_true} shows an example of fitted risk scores (random forest model) against true risks (generated from a generalised linear model). Panel~\ref{fig:rho_eq_by_cohort} shows values of $\rho_{eq}$ by age- and medical history-defined cohorts. Panel~\ref{fig:intervention_continuous} shows the effect of an intervention on a continuous covariate (diet, alcohol; both in $[0,10]$). The post-intervention value is uniformly distributed with a range dependent on the initial value (3 in this case) and the difference $\rho-\rho_{eq}$. Panel~\ref{fig:intervention_binary} shows the effect of an intervention on a binary covariate (diet, alcohol; both in $[0,10]$).} 
\label{supp_fig:simulation_details}
\end{figure}

\end{bibunit}

\end{document}